\documentclass[iop,twocolumn]{aastex631}
\usepackage{CJK}
\usepackage{mathrsfs}
\usepackage{amsmath}
\usepackage{float}
\usepackage{amstext}
\usepackage{soul}

\hypersetup{
   colorlinks,
   linkcolor={blue!88!black!80},
   citecolor={blue!88!black!80},
   urlcolor={blue!88!black!80}}

\newcommand {\hb}{\ifmmode {\rm H}\beta \else H$\beta$\fi}
\newcommand {\ha}{\ifmmode {\rm H}\alpha \else H$\alpha$\fi}

\begin{document}

\begin{CJK}{UTF8}{gbsn}

\title{AT2021aeuk: A Repeating Partial Tidal Disruption Event Candidate in a Narrow-line Seyfert 1 Galaxy}

\correspondingauthor{Jingbo Sun, Hengxiao Guo }
\email{Emails: sunjingbo@shao.ac.cn (JBS),\\ hengxiaoguo@gmail.com (HXG)} 

\author[0000-0003-3472-4392]{Jingbo Sun}
\affiliation{Shanghai Astronomical Observatory, Chinese Academy of Sciences, 80 Nandan Road, Shanghai 200030, People's Republic of China}
\affiliation{University of Chinese Academy of Sciences, 19A Yuquan Road, 100049, Beijing, People's Republic of China}
\author[0000-0001-8416-7059]{Hengxiao Guo}
\author[0000-0002-4455-6946]{Minfeng Gu}
\author[0000-0002-7329-9344]{Ya-Ping Li}
\author[0000-0001-5650-6770]{Yongjun Chen}

\affiliation{Shanghai Astronomical Observatory, Chinese Academy of Sciences, 80 Nandan Road, Shanghai 200030, People's Republic of China}

\author[0000-0002-9280-1184]{D. Gonz\'{a}lez-Buitrago}
\affiliation{Universidad Nacional Aut\'onoma de M\'exico, Instituto de Astronom\'ia, AP 106,  Ensenada 22860, BC, M\'exico}
\affiliation{Department of Physics and Astronomy, 4129 Frederick Reines 
Hall, University of California, Irvine, CA, 92697-4575, USA}

\author[0000-0003-4156-3793]{Jian-Guo Wang}
\author[0000-0003-3823-3419]{Sha-Sha Li}
\author[0000-0002-1530-2680]{Hai-Cheng Feng}
\author[0000-0002-6809-9575]{Dingrong Xiong}
\affiliation{Yunnan Observatories, Chinese Academy of Sciences, Kunming 650011, Yunnan, People's Republic of China}
\affiliation{Key Laboratory for the Structure and Evolution of Celestial Objects, Chinese Academy of Sciences, Kunming 650011, Yunnan, People's Republic of China}

\author[0000-0003-3207-5237]{Yanan Wang}
\affiliation{National Astronomical Observatories, Chinese Academy of Sciences, 20A Datun Road, Beijing 100101, People's Republic of China}

\author[0000-0003-4671-1740]{Qi Yuan}
\affiliation{Changchun Observatory, National Astronomical Observatories, Chinese Academy of Sciences, Changchun 130117, People's Republic of China}

\author[0000-0002-8402-3722]{Jun-jie Jin}
\author[0000-0003-1702-4917]{Wenda Zhang}
\affiliation{National Astronomical Observatories, Chinese Academy of Sciences, 20A Datun Road, Beijing 100101, People's Republic of China}

\author[0000-0001-6858-1006]{Hongping Deng}
\affiliation{Shanghai Astronomical Observatory, Chinese Academy of Sciences, 80 Nandan Road, Shanghai 200030, People's Republic of China}
\author[0009-0006-5706-0364]{Minghao Zhang}
\affiliation{National Astronomical Observatories, Chinese Academy of Sciences, 20A Datun Road, Beijing 100101, People's Republic of China}



\begin{abstract}
A black hole (BH) can tear apart a star that ventures within its tidal radius, producing a luminous flare as the stellar debris falls back, known as a tidal disruption event (TDE). While TDEs in quiescent galaxies are relatively well understood, identifying TDEs in active galactic nuclei (AGN) still remains a significant challenge. We present the discovery of AT2021aeuk, a transient exhibiting dual flares within around three years in a narrow-line Seyfert 1 galaxy. Multi-wavelength observations triggered during the second flare in 2023 revealed an extraordinary X-ray V-shaped light curve, strongly anti-correlated with the optical light curve and accompanied by a lag of $\sim$40 days. This behavior is inconsistent with both supernova and pure AGN origins. In addition, a new broad component emerges in the Balmer lines during the second flare, showing a clear reverberation signal to the continuum variation. We propose that the dual-flare may be linked to a repeating partial TDE (rpTDE), where the second flare results from a collision between the TDE stream and the inner accretion disk, triggering an optical flare while simultaneously partially destroying the X-ray corona. However, other mechanisms, such as a stellar-mass BH (sBH) merger within an accretion disk, could produce similar phenomena, which we cannot entirely rule out. The Vera C. Rubin Observatory will be a powerful tool for further investigating the nature of such events in the future.

\end{abstract}



\section{Introduction} \label{sec:intro}

At the center of galaxies, supermassive black holes (SMBHs) play a crucial role in galaxy formation and evolution \citep[e.g.,][]{Kormendy2013}. A small fraction of SMBHs are active, surrounded by an accretion disk that powers multi-wavelength emissions, known as active galactic nuclei \citep[AGNs, ][]{Reynolds1997, Sanders1988}. They typically exhibit stochastic variability across the electromagnetic spectrum \citep{Ulrich1997}, with UV/optical continuum emission varying by approximately 0.2 mag over timescales of months to years \citep{Sesar2007, MacLeod2012}. Although rare, some AGNs display extreme variability (e.g., $>$ 1 mag), which lies at the tail end of a broader distribution of AGN properties \citep{Rumbaugh2018, Guo2020a, Ren2022}. This extreme population provides a unique opportunity to gain deeper insights into the accretion disk and the underlying mechanisms driving variability \citep{Wang2023a, Graham2020}.

In the past decade, one of the most significant discoveries related to SMBHs has been the identification of tidal disruption events (TDEs) showing remarkable UV/optical flares \citep[][see review]{Gezari2021}. TDEs occur when a star crosses the tidal radius of an SMBH and is torn apart by its gravitational forces \citep{Rees1988}. These events typically exhibit a rapid rise followed by a slow, power-law decay in their UV/optical light curves \citep{vanVelzen2011, Gezari2012}, regulated by the mass fallback rate \citep{Phinney1989}. Thanks to advancements in time-domain surveys and well-organized follow-up observations, more characteristic features have been clearly identified, including constant temperature evolution \citep{Chornock2014, Gezari2015}, very broad emission lines with widths exceeding 10000 $\rm km\ s^{-1}$ \citep{vanVelzen2020}, coronal and high-ionization lines \citep{Hammerstein2023b}, and delayed or absent X-ray emission \citep{Guolo2024}. Notably, the constant color or single blackbody temperature evolution observed during the decay phase is a distinguishing feature that sets TDEs apart from AGNs and supernovae (SN) \citep{vanVelzen2021}. The growing sample of known TDEs is enabling the study of their statistical properties \citep{vanVelzen2021, Hammerstein2023b, Yao2023}. However, a definitive criterion for distinguishing TDEs from other transient events remains elusive \citep{Zabludoff2021}, and their identification continues to be largely empirical, relying on previously identified characteristics that may introduce selection biases \citep{Gezari2021}.

The overall TDE paradigm and the origins of multi-wavelength emissions are still under debate, with two broad scenarios emerging. The first is the reprocessing scenario, where a compact accretion disk forms rapidly after the stellar debris falls back. X-ray photons from the disk are subsequently reprocessed into UV/optical emission by an optically thick envelope or outflows \citep{Loeb1997, Ulmer1999, Strubbe2009, Coughlin2014, Metzger2016, Roth2016, Dai2018, Metzger2022,Thomsen2022}. The second scenario attributes the UV/optical emission to shocks generated by collisions between the fallback and outward-moving debris streams, while the X-ray emission arises from the delayed formation of an accretion disk \citep{Piran2015, Dai2015, Shiokawa2015, Jiang2016, Alexander2016, Ryu2023, Guo2025}.

TDEs can take place in both quiescent and active galaxies. Theoretical studies suggest a potentially higher occurrence rate in AGNs compared to inactive galaxies \citep{Karas2007, Just2012, Kennedy2016}. Given that the TDE rate in quiescent galaxies is estimated at around $10^{-4}$ to $10^{-5}$ yr$^{-1}$ galaxy$^{-1}$ \citep{vanVelzen2018, Yao2023}, and with nearly one million known AGNs \citep{Flesch2023}, at least a handful of TDEs should be expected among these AGNs. However, the observable characteristics of TDEs in AGNs remain undefined both in observations and theories. On one hand, it is unclear how the flare from gas accretion in an AGN might differ from a TDE in a quiescent galaxy. On the other hand, the evolution of the TDE fallback stream could be affected by the pre-existing accretion flow \citep{Kathirgamaraju2017}, potentially leading to unique behaviors within an AGN. These interactions remain largely unexplored, mainly due to the difficulties in identifying TDEs in AGNs. 


Recently, several candidate TDEs in known AGNs have been extensively studied through multi-wavelength observations, such as PS10adi \citep{Kankare2017}, PS16dtm \citep{Blanchard2017, Petrushevska2023}, 1ES 1927+654 \citep{Trakhtenbrot2019b, Ricci2020}, AT2018dyk \citep{Frederick2019, Huang2023a}, OGLE16aaa \citep{Wyrzykowski2017, Kajava2020}, F01004-2237 \citep{Tadhunter2017, Sun2024}, and AT2019fdr and AT202hle \citep{Frederick2021}. The features of these TDE candidates in AGNs differ from those of TDEs in quiescent galaxies and also vary significantly among themselves, suggesting a more complex scenario for TDE in AGN. Furthermore, theories developed for both AGNs and TDEs can account for a broad range of behaviors in flaring transients \citep{Moriya2017, Shu2020}, making it even more challenging to establish universal criteria for identifying TDEs in AGNs.

\begin{figure*}
    \centering
    \includegraphics[width=\textwidth]{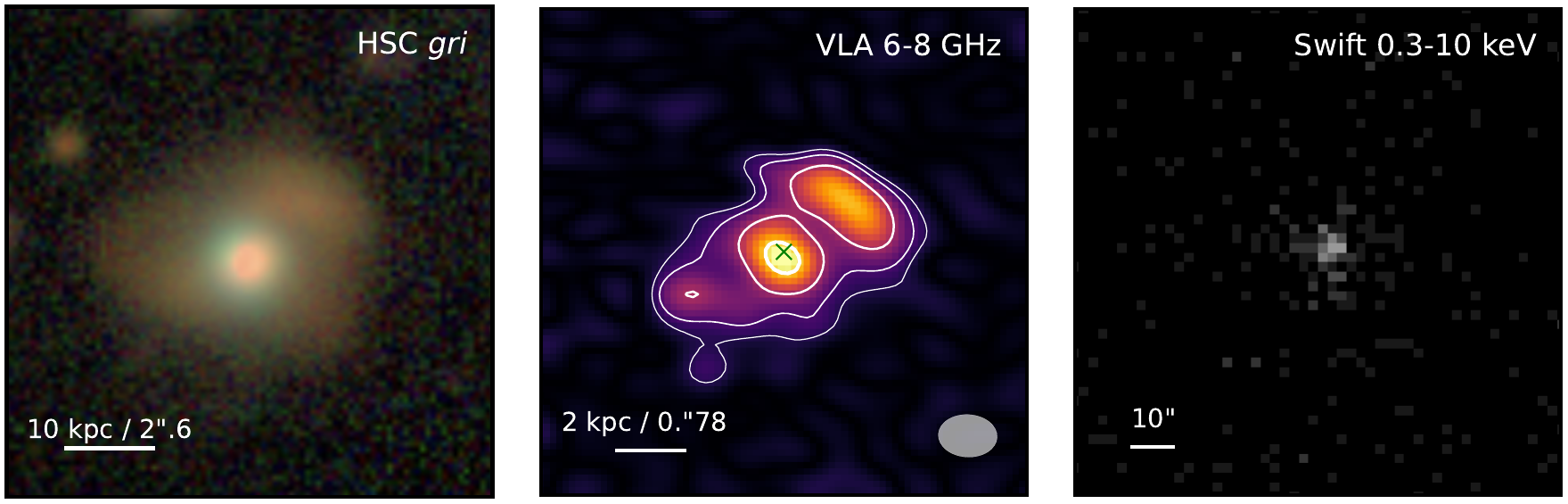}
     \caption{The HSC optical, VLA radio, and Swift X-ray images for AT2021aeuk. \textit{Left panel}: HSC $gri$ composite image with a 15\arcsec $\times$ 15\arcsec\ field of view (FOV) observed before the flares of AT2021aeuk. \textit{Middle panel}: VLA $6-8$ GHz image with a 4\arcsec $\times$ 4\arcsec\ FOV using A-configuration during the second flare. The gray ellipse and green cross indicate the beam size and GAIA optical center, respectively. Diffuse components are discovered on both sides separated 2 kpc from the center. \textit{Right panel}: Stacked X-ray image with an FOV of 60\arcsec $\times$ 60\arcsec\ from all the Swift XRT observations, most of which are obtained during the second flare.}
    \label{fig:3image}
\end{figure*}

In certain cases, a TDE can trigger multiple flares if the star is only partially disrupted and remains bound to the SMBH, a phenomenon known as a repeating partial TDE (rpTDE). Among TDEs identified in quiescent galaxies, recurring flares in the UV/optical have been observed, displaying typical TDE characteristics \citep{Wevers2023, Somalwar2023a, Lin2024}. It is worth noting that AT2022dbl exhibited two flares with similar spectral features \citep{Lin2024}, possibly indicating that both flares originated from the same star. RpTDE candidates have also been reported in AGNs, such as ASASSN-14ko \citep{Payne2021, Payne2022, Payne2023, Huang2023b}, AT2019aalc \citep{Veres2024}, and F01004-2237 \citep{Tadhunter2017, Sun2024}. These rpTDE candidates provide insights into stellar trajectories and serve as testbeds to investigate the potential differences between partially and fully disrupted TDEs.

Here, we report the discovery of repeating flares in AT2021aeuk, which experienced its first flare in 2019 and a second one in 2023. Its first flare was noticed at the end of 2022 when we were compiling archival light curves of the NLS1 sample \citep{Foschini2015} from the Zwicky Transient Facility \citep[ZTF,][]{Bellm2019}. Occasionally, a ZTF alert \citep{Forster2021} reported a second flare on 2022 December 23, prompting us to trigger multi-wavelength observations during the second flare.

This article is organized as follows: multi-wavelength observations and results are described in \S \ref{sec:observation} and \S \ref{sec:results}, respectively. We discusses the possible mechanisms that triggered AT2021aeuk in \S \ref{diss:flare}. Finally, we summarize our main findings in \S \ref{sec:conclusion}. Throughout this paper, we adopt a flat $\Lambda$CDM cosmology with $H_{0}=70\ \rm km\ s^{-1}\ Mpc^{-1}$ and $\Omega_{\rm m}$ = 0.3.

\section{Observations} \label{sec:observation}
\subsection{Pre-outburst State of AT2021aeuk}\label{sec:host}

\begin{figure*}
    \centering
    \includegraphics[width=0.95\textwidth]{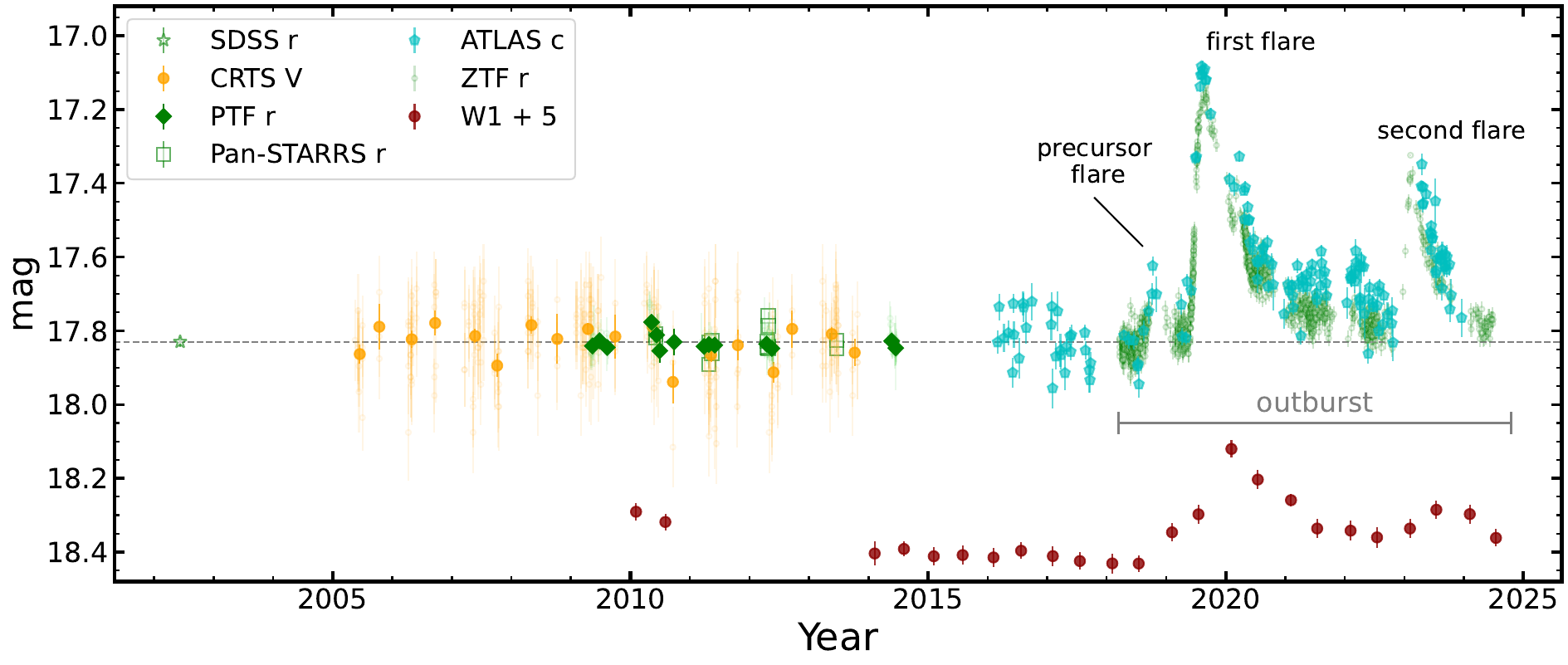}
    \caption{Long-term light curves of AT2021aeuk. CRTS data are seasonally binned for clarity. No dramatic variability was detected during the 15-year baseline prior to the outburst.}
    \label{fig:long_lc}
\end{figure*}

AT2021aeuk coincides with the center of SDSS J161259.83+421940.3, a narrow-line Seyfert 1 (NLS1) galaxy at redshift $z = 0.2336$ \citep{Foschini2015}, first identified from Sloan Digital Sky Survey (SDSS) spectroscopic observations in 2003 \citep{York2000}. Evidence of nuclear activity before the outburst is provided by WISE colors \citep{Stern2012}, narrow line ratios \citep{Baldwin1981}, and strong X-ray emission. Using the pre-outburst SDSS spectrum, \cite{Foschini2015} estimated the black hole (BH) mass to be $8.8 \times 10^{6}\ M_{\odot}$ based on the empirical H$\beta$ radius-luminosity relation. To estimate the bolometric luminosity, we integrated the spectral energy distribution (SED) from 1 $\mu$m to 8 keV, yielding $L_{\rm bol} = 5.3 \times 10^{44}\ \rm erg\ s^{-1}$, resulting in an Eddington ratio of $\sim0.5$.


Figure~\ref{fig:3image} displays the Hyper Suprime-Cam (HSC) image \citep{Aihara2018}, along with the radio image from Karl G. Jansky Very Large Array (VLA) observations and the X-ray image from the Neil Gehrels Swift Observatory. The HSC image was observed by the 8.2~meter Subaru telescope prior to the outburst \citep{Aihara2022}. Although the displayed VLA image at $6-8$ GHz was taken during the second flare, we consider it representative of the pre-outburst state, given no significant radio variation during the flare (see \S \ref{sec:radio}). In addition, we present the all epoch stacked X-ray images, as the luminosities before and after the outburst are comparable (see \S \ref{sec:x-ray}).



We compiled the long-term optical light curve using photometric data from SDSS \citep{York2000}, the Catalina Real-time Transient Survey \citep[CRTS,][]{Drake2009}, the Panoramic Survey Telescope and Rapid Response System \citep[Pan-STARRS,][]{Chambers2016}, the Palomar Transient Factory \citep[PTF,][]{Law2009}, the Asteroid Terrestrial-impact Last Alert System \citep[ATLAS,][]{Tonry2018}, and ZTF. Following \cite{pycali2014}, the light curves from different surveys were aligned to a common baseline, with the ZTF $r$-band serving as the reference. Meanwhile, the ATLAS data were scaled to match both the flux level and the variability amplitude of ZTF. As shown in Figure~\ref{fig:long_lc}, the long-term light curve displays weak variability of less than 0.1 mag over a 15-year period preceding the outburst. During the outburst, two major flares and a minor precursor flare were observed, with the latter being clearly captured by ATLAS. Since the ZTF and ATLAS light curves overlap, we focus on the ZTF data in our subsequent analyses due to its larger telescope aperture and superior photometric accuracy. These optical flares from AT2021aeuk have also been noticed by \cite{Bao2023}.

\subsection{Optical Photometric Observations}
We triggered photometric observations using the Lijiang 2.4-meter telescope (LJT) in the $gri$ bands with a daily cadence near the peak of the second flare (PI: J. Wang). The data were reduced using \texttt{PyRAF} following standard procedures, and aperture photometry was performed to extract the flux. The resulting light curve is consistent with the contemporaneous ZTF observations, without the need for further calibration.

\subsection{Swift UV and X-ray Observations}
Three observations were retrieved from the archival data of the Swift Observatory \citep{Gehrels2004}. The first detection occurred before the first flare, while the other two detections took place during the quiescent period between the first and second flares, somehow entirely missing the flare phases. All three detections showed similar luminosities of approximately $7.8 \times 10^{43} \ \rm erg \ s^{-1}$. Our target-of-opportunity observations (Program IDs 18102, 18408, 18545, 19097, 19392, 19719, 19930, 20347; PI: J. Sun) were triggered promptly after the onset of the second flare, with an exposure time of 2000 s per visit, shared across all requested filters. Particularly, daily cadence observations were conducted near the peak of the second flare using four Swift UV filters ($U$, $UVW1$, $UVM2$, $UVW2$). However, the allocated exposure time was insufficient to produce high-quality light curves across all four bands to estimate the lags. Thus, we modified our observational strategy and concentrated on the $UVW1$ band for subsequent observations during the second flare.

We processed the XRT data using the online tools provided by the UK Swift Science Data Center, which uses {\tt XRTPIPELINE} and {\tt XRTPRODUCTS} to extract the source with a circular region of 20 pixels, and estimate the background using an annulus region from 50 to 60 pixels. Combining all the data with a total of 111.7~ks exposure, we get a point-like X-ray source at the location of AT2021aeuk. After determining the detection centroid, the X-ray observations are binned by counts to extract the light curve. The number of counts per bin is 15, with a minimum value of 5. Since the X-ray spectra are similar across all observations, the counts-to-flux conversion factor for $0.3 -10$ keV is estimated to be $2.88 \times 10^{-11}\ \rm erg\ cm^{-2}\ ct^{-1}$ from the stacked spectrum from all X-ray observations, which can be adequately fitted by a power-law model, with a photon index $\Gamma$ of $2.46 \pm 0.07$. The UVOT photometry is performed by \texttt{uvotsource} task, using a 5\arcsec\ aperture on the optical center, which is sufficient to enclose the source. Then a nearby 40\arcsec\ radius aperture is adopted to estimate the background level.

\subsection{WISE Mid-infrared Archival Observations}
Wide-field Infrared Survey Explorer \citep[WISE,][]{Wright2010} scanned the entire sky in $W1$ and $W2$ bands with a half-year cadence from the launch on 2009 December 14 until the completion on 2024 August 1, apart from a hibernation period from 2011 February to 2013 December. We retrieved the mid-infrared (MIR) light curves from the ALLWISE catalog and the final data release of NEOWISE-R \citep{Mainzer2014}, which includes observations up to the last visit on 2024 July 18. Following the official recommendations, we removed photometric measurements affected by poor-quality frames, charged particle hits, scattered moonlight, or artifacts. The exposures from each semiannual visit were binned into a single data point to enhance the signal-to-noise ratio (SNR).

\begin{figure*}[ht!]
\begin{center}
    \includegraphics[width=0.9\linewidth]{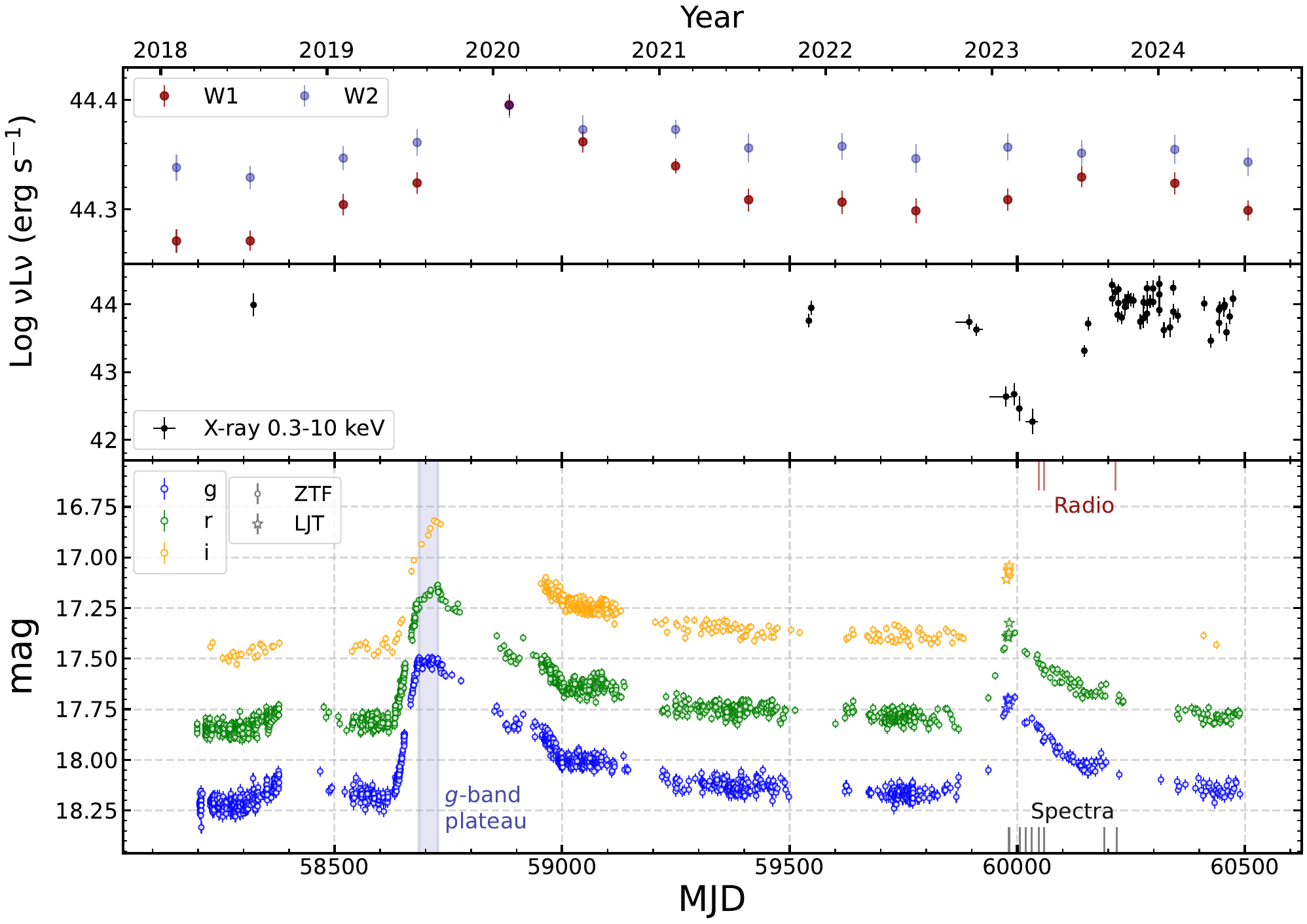}
    \caption{\textit{Top panel}: The MIR light curves, exhibiting significant echos to the optical flares. \textit{Middle panel}: The X-ray light curve, with a nearly inverse evolution pattern to the optical light curve of the second flare. \textit{Bottom panel}: ZTF (circles) and LJT (stars) light curves of the flares. The blue shadow indicates the period of the $g-$band plateau phase. The spectroscopic and radio observations are indicated by the black and red short lines, respectively. \label{fig:lc}}
\end{center}
\end{figure*}

\subsection{Radio Observations}\label{sec:radio}
We obtained radio observations of AT2021aeuk at C ($4-8$ GHz), X ($8-12$ GHz), and Ku ($12-16$ GHz) bands during the second flare using VLA. The first two observations (Program ID 23A-390; PI: M. Gu) were conducted on 2023 April 14 (+44 days in the rest frame, relative to the peak of the second flare on 2023 February 17 and hereafter) and April 25 (+53 days) with the B-configuration. Subsequent observations with A-configuration (Program ID 23B-306; PI: J. Sun) at the same frequency range were conducted on 2023 September 29 (+182 days). The data were reduced using standard procedures in the Common Astronomy Software Applications (CASA) package \citep{McMullin2007}. Two epoch observations with B-configuration did not reveal significant flux variations and then were combined to make images with improved SNRs. The radio observations at 5 GHz with A-configuration revealed extended radio emission near the nuclear region, i.e., two diffuse emission regions located 2 kpc on either side of the nucleus. The collected archival radio flux at other wavebands at pre-outburst time are well matched with the power-law fit from our VLA radio flux measurements during the second flare. This suggests the VLA radio emission is likely from the AGN itself, rather than the fresh emission caused by transient flares. Therefore, we do not discuss the radio properties in the following text. The details of the radio information including the morphology and spectrum are given in Appendix \ref{app:radio}.

\begin{deluxetable*}{lccccccc}
\tablecaption{{The fitting results of $g$-band light curve} }
\tablehead{\colhead{Flare} & \colhead{$L_{\rm peak}$}  & \colhead{$\sigma$}  & \colhead{$t_{\rm peak}$} & \colhead{$t_{0}$} & \colhead{$p$}& \colhead{$\Delta t_{\mathrm{plateau}}$}\\
\colhead{} & (erg $\rm s^{-1}$)&(day)&(MJD)&(day)& &(day)}
\startdata
    first flare      & 1.7 $\times$ $10^{44}$ & 25.7 & 58690.0 & 323.9 & $-2.2$ & 40.2 \\
    second flare     & 1.1 $\times$ $10^{44}$ & 31.9 & 59992.1 & 355.9 & $-3.4$ &      \\
    \hline
    precursor flare & 3.2 $\times$ $10^{43}$ & 63.9 & 58445.4 &       &      &      \\
    bump58932            & 2.7 $\times$ $10^{43}$ & 26.3 & 58932.3 &       &      &      \\
    bump59099            & 9.6 $\times$ $10^{42}$ & 25.2 & 59099.3 &       &      &      \\
    bump60199            & 1.5 $\times$ $10^{43}$ & 16.9 & 60198.6 &       &      & \\
\enddata
\tablecomments{$\sigma$ is the variance of Gaussian rise, $t_{0}$ is the power-law normalization, and $p$ is the power-law index. 
\label{tab:lc_fit}}
\end{deluxetable*}   

\subsection{Spectroscopic Observations}
During the second flare, we obtained seven epochs of spectroscopic observations: three epochs (+22, +32, and +45 days) using the 10.4-meter Gran Telescopio CANARIAS (GTC) through Director's Discretionary Time (DDT) (Program ID: GTC2023-208; PI: D. Gonz\'{a}lez-Buitrago), three epochs ($-$8, +11, and +161 days) from the 2.4-meter Lijiang Telescope (LJT) (PI: J. Wang), and one epoch (+180 days) from the 2.16-meter Xinglong Telescope (XLT) (PI: J. Sun).

The GTC observations are conducted using the OSIRIS+ with 1000R and 1000B grism. Each grism takes three 900s exposures with a 0.6\arcsec\ slit under the seeing of $\sim 0.9\arcsec$, yielding an observed wavelength coverage of 3600$-$13000\ \AA\ and a resolution of $R\sim 1100$. For the LJT, we use the grism 8 and 14 of Yunnan Faint Object Spectrograph and Camera (YFOSC) to obtain the red and blue sides of the spectrum, respectively. Each grism is exposed for 30 min, providing a dispersion of around 1.8 \AA\ pixel$^{-1}$ in the blue side and 1.5 \AA\ pixel$^{-1}$ for the red side, totally covering from  3600$-$9800~\AA. The XLT spectrum is exposed for 1 hr using the grism 8 of the Beijing Faint Object Spectrograph and Camera (BFOSC), with a dispersion of 1.7 \AA\ pixel$^{-1}$, covering from 4000$-$8800~\AA. All spectra were reduced and calibrated using \texttt{pypeit} \citep{Prochaska2020} or \texttt{PyRAF}. Additional flux calibration was performed using the [\ion{O}{3}] $\lambda$5007 line, assuming it remained constant throughout the monitoring. 


Following \cite{Shen2011}, we performed local spectral fitting in [6000, 7300]\AA\ range for H$\alpha$ and [4300, 5500]\AA\ range for H$\beta$ using \texttt{PyQSOFit} \citep{Guo2018}. The local continuum was modeled using a power-law and an iron template \citep{Boroson1992}. After subtracting the continuum, two Gaussians were used to model each line of the [\ion{O}{3}]$\lambda \lambda$4959,5007 and [\ion{S}{2}]$\lambda \lambda$6718,6732 emission lines, representing a narrow and an intermediate component \citep{Greene2005}. The widths and offsets of each narrow and intermediate component were tied together in each complex, and the ratio between [\ion{S}{2}] doublets was fixed at 1.2:1. An extra broad component was included for both H$\alpha$ and H$\beta$. The emission line fitting results are presented in Appendix \ref{app:emission_line}, which demonstrates that a new broad component emerged during the second flare.

The obtained flux ratio of total $\rm H\beta$ to [\ion{O}{3}]$\lambda$5007 is approximately 1:2 based on the spectral decomposition of SDSS spectrum. Additionally, the fitted FWHMs of the intermediate components of H$\alpha$ and H$\beta$ are 1100 km~s$^{-1}$ and 1400 km~s$^{-1}$, respectively, both satisfying the NLS1 classification criteria of [\ion{O}{3}]$/\rm H\beta < 3$ and $\rm FWHM < 2000\ km\ s^{-1}$ \citep{Osterbrock1985}. Furthermore, the FWHM ratio between $\rm H\beta$ and $\rm H\alpha$ was around 0.79, consistent with the typical FWHM(H$\alpha$)/FWHM(H$\beta$) relation observed in NLS1s \citep{Paliya2024}.

\section{Results} \label{sec:results}
In this section, we examine the UV/optical, X-ray, and MIR light curves, as well as the associated spectral features. Given the similarities between the UV/optical flares observed in our study and those in TDEs, we use TDEs as a reference in our following discussions.

\begin{figure}
\hspace{-1cm}
    \centering
    \includegraphics[width=0.5\textwidth]{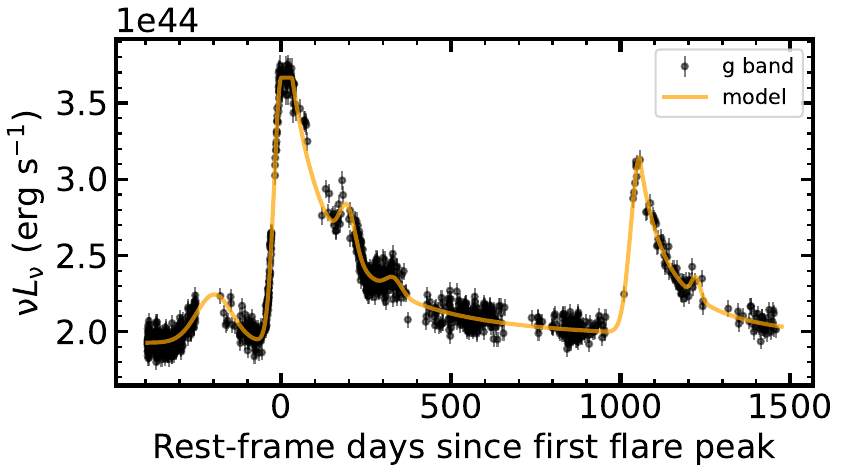}
    \caption{The $g$-band light curve, modeled using various components: Gaussian profiles represent the precursor flare, rising phases of the two major flares, and three additional bumps; power-law functions describe the declines; and a plateau illustrates the peak.}
    \label{fig:lc_fit}
\end{figure}

\subsection{UV and Optical Flares} \label{sec:lc}
\subsubsection{Comparison with TDEs in Quiescent Galaxies}
As illustrated in Figure \ref{fig:lc}, AT2021aeuk displays two significant flares approximately three years apart, each lasting about one year. These flares are characterized by a rapid increase followed by a gradual decline, reminiscent of the optical flares observed in TDEs. To quantify the amplitude and evolution of these flares, we fit the optical $g$-band light curve using a parametric model. This model includes Gaussian profiles for the precursor flare, rise phases for the two primary flares and bumps, a power-law for the decline, and a plateau to represent the flat peak. The fitting results are shown in Figure~\ref{fig:lc_fit}, and the fitted parameters are listed in Table \ref{tab:lc_fit}. The luminosity ratios for the first, second, and precursor flares are $1:0.65:0.2$, respectively. The power-law index of the decline for the first flare is $\alpha = -2.2 \pm 0.16$, consistent with the theoretical decline rate of $\propto t^{-9/4}$ for a partial TDE \citep{Coughlin2019}. The slope of the second flare is relatively steeper with an index of $\alpha = -3.4 \pm 0.2$, probably influenced by the incomplete data on the declining tail. In general, the overall shape and evolution of the two flares are similar.

\begin{figure}
\begin{center}
    \includegraphics[width=0.45\textwidth]{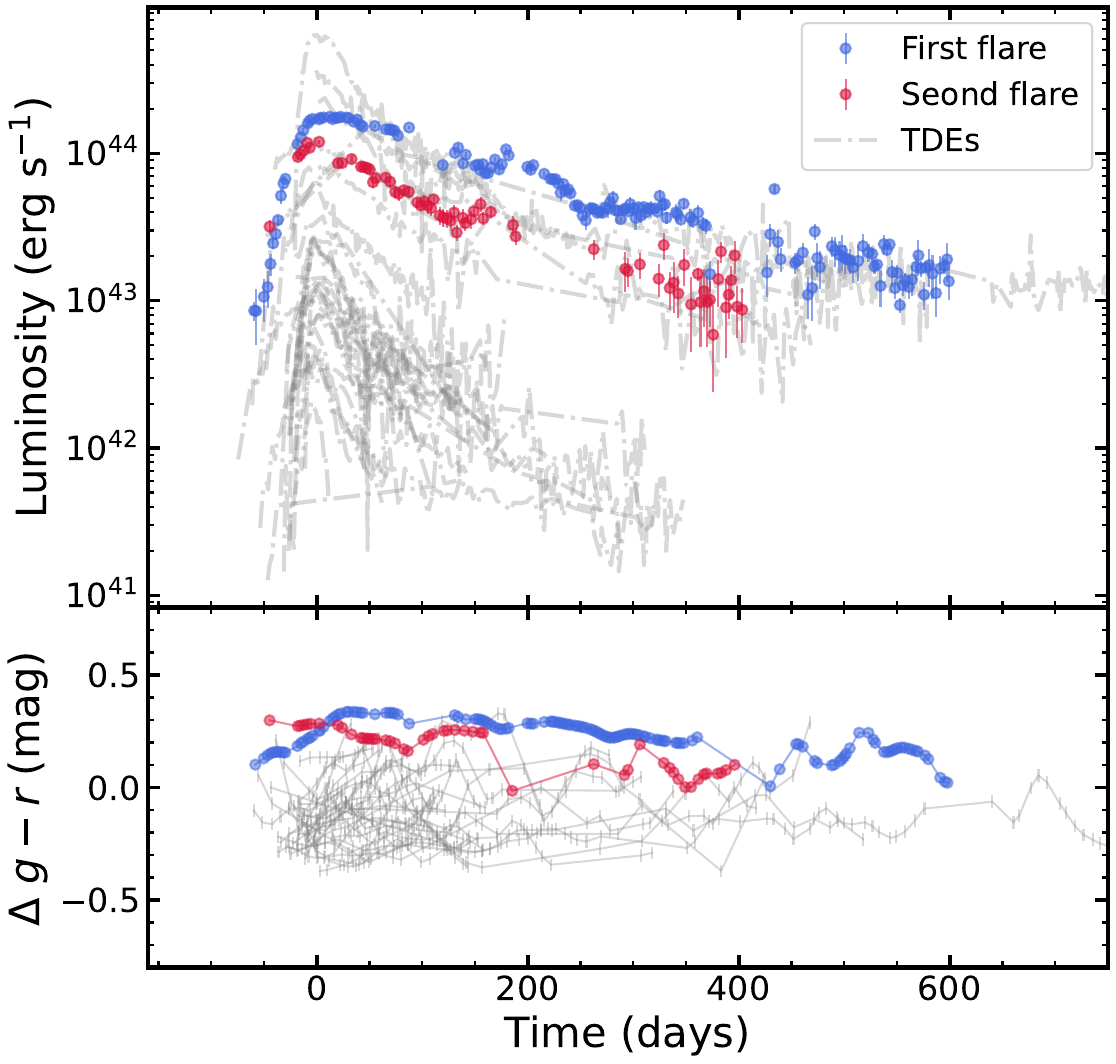}
    \caption{\textit{Top panel}: pre-outburst subtracted $g$-band light curves of the two flares of AT2021aeuk in AGN and other identified TDEs in quiescent galaxies \citep{Yao2023}. \textit{Bottom panel:} the $\Delta g-r$ color evolution during the flares. }
    \label{fig:compare_lc}
\end{center}
\end{figure}

\begin{figure}
    \centering
    \includegraphics[width=0.45\textwidth]{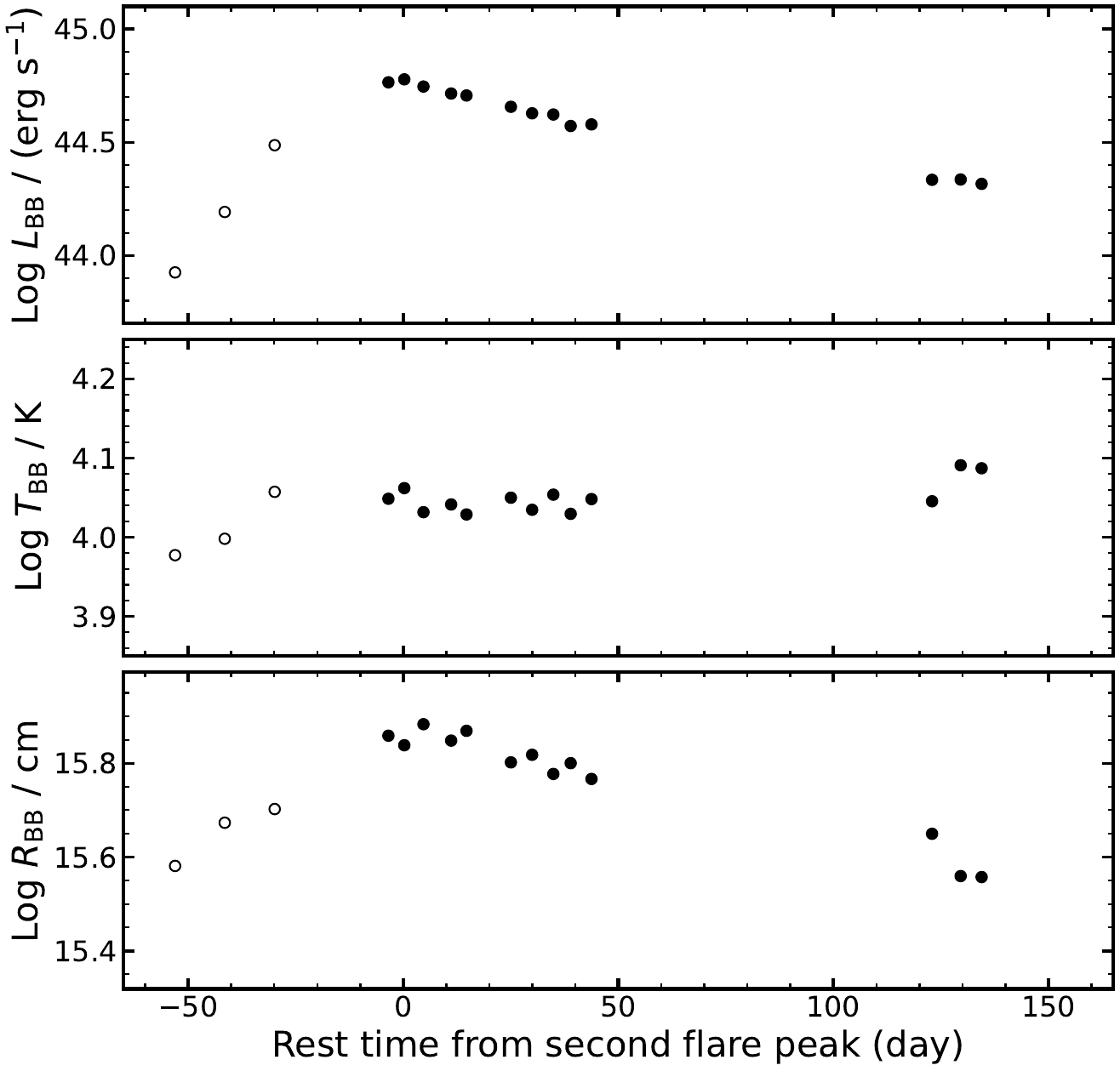}
    \caption{Evolution of the blackbody luminosity, temperature, and radius during the second flare of AT2021aeuk, using non-parametric fitting methods. Each epoch's fitting is based on $UVW2$, $UVW1$, and interpolated $g$- and $r$-band luminosities derived from the fitted light curve. Hollow points represent epochs lacking $UVW2$ observations, where fittings are performed using the remaining three bands. }
    \label{fig:TBB}
\end{figure}

We present a comparison of the light curves for the two flares of AT2021aeuk with those of identified TDEs in quiescent galaxies \citep{Yao2023} in Figure~\ref{fig:compare_lc}. The evolution of AT2021aeuk's light curves during these events parallels the brightest, longest-lasting TDEs in Yao's sample. Notably, the $g-r$ colors of the two flares are slightly redder around the peak but gradually transition to bluer values, eventually aligning with those observed in TDEs in quiescent galaxies. This shift in color, indicative of a modest increase in temperature during the decline phase (see also Figure \ref{fig:TBB}), supports the TDE hypothesis and disagree with an AGN or SN origin for these flares. Further comparative analysis of light curve parameters between AT2021aeuk and known TDEs, as well as other nuclear transients, is detailed in Appendix \ref{app:compare}. Our findings suggest that the flares from AT2021aeuk exhibit characteristics intermediate between typical TDEs and other nuclear events.

\begin{figure}
    \centering
    \includegraphics[width=0.45\textwidth]{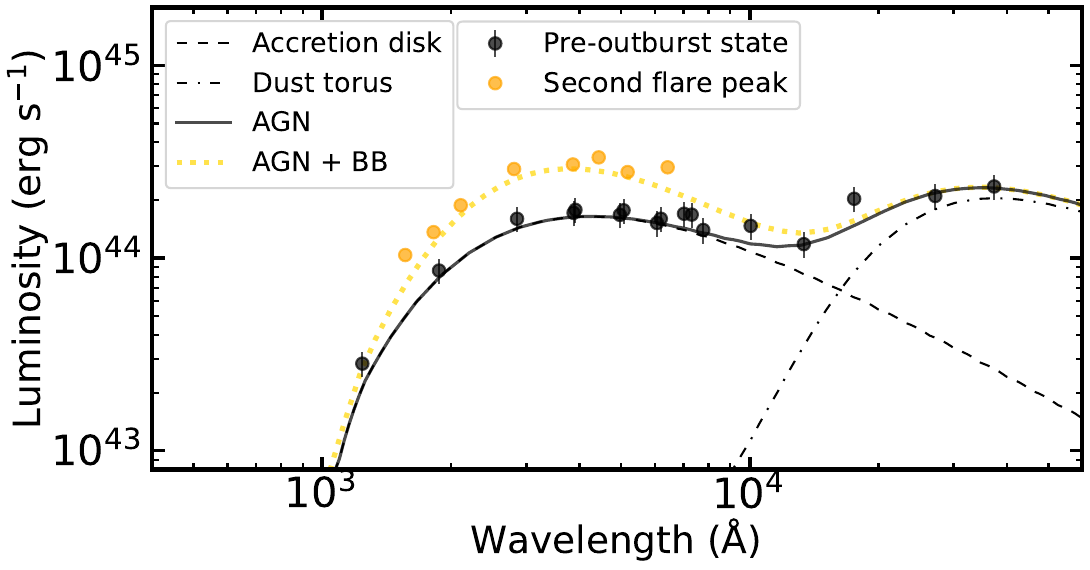}
    \caption{SEDs for the pre-outburst state (black) and the second flare (orange). The first flare's SED is decomposed into accretion disk and dust torus components, while the SED of the second flare peak incorporates an additional blackbody component into the pre-outburst SED. Photometric data from UVOT, binned within $\pm 15$ days of the second flare peak, represent the peak luminosity. All SEDs are modeled using X-CIGALE \citep{Yang2020}.}
    \label{fig:sed}
\end{figure}

To investigate the SED evolution, we initially modeled the pre-outburst SED using archival multi-wavelength data via X-CIGALE \citep{Boquien2019, Yang2020}. After removing the pre-outburst luminosities, the differential flare SED closely resembled a single blackbody with a temperature of $10^4$ K. The results of the SED fittings for both the pre-outburst stage and the second flare peak are depicted in Figure \ref{fig:sed}. Following this method each epoch is performed blackbody fits independently using $UVW2$, $UVW1$, $g$, and $r$ photometries. The evolution of the fitted blackbody parameters, shown in Figure \ref{fig:TBB}, maintains a nearly constant temperature post-peak, supporting the classification of the second flare as a potential TDE. However, caution is warranted when using single blackbody fits with limited wavelength coverage, as the actual SED of a TDE may not conform to a single blackbody model. \citep[e.g.,][]{Leloudas2019}. The comparison of the fitted parameters with other TDE candidates and nuclear transients is presented in Appendix \ref{app:compare}.

\subsubsection{Three Unique Features in Light Curves}
Our target exhibits three unique features in the UV/optical light curves, including a plateau at the peak of $g$-band light curve, a precursor flare preceding the two major flares, and a UV light curve that returns to a state even lower than that in pre-outburst stage (or a UV dip).

The first feature is a plateau phase lasting about 40 days observed in the $g$-band light curve at the first flare peak. Simultaneously, the $r$ and $i$ bands turn into a slow-increasing trend coinciding with the $g$-band plateau, with longer wavelength band having steeper increasing trend, supporting the feature is real rather than any artifacts. The exact cause is unclear, and the delayed increase observed in the $r$ and $i$ bands might be attributed to contamination from the $\rm H \beta$ and $\rm H \alpha$ lines, respectively, or to extra reprocessing within the $ri$ continuum emitting region. The extra increase, around 0.1 mag in $r$ band and 0.2 mag in $i$ band, aligns with the larger variability in $\rm H\alpha$ compared to $\rm H\beta$. This $g$-band plateau is notably stable, with a standard deviation of only $\sim$0.015 mag. Such a flat peak is rarely observed in other transients. For example, \cite{Blanchard2017} reported a flat peak lasting for $\sim$ 80 days from UV to optical bands in another TDE in AGN candidate PS16dtm. However, its flat peak is likely mimicked by two close luminosity peaks. We suggest that the plateau phase of our target might result from reaching the Eddington limit \citep{Loeb1997}, or from the equilibrium established between dissipation, inflow, and cooling during the interaction between the TDE fallback stream and the accretion disk \citep{Chan2020}.


The second feature is a precursor flare, occurring approximately 200 days before the first major flare and reaching about 20\% of the luminosity of the first flare and exhibits a nearly symmetric evolution, though with sparse sampling as shown in Figure \ref{fig:long_lc}. Unlike the early bumps that occur during the rising phase of some TDEs \citep{Charalampopoulos2023,Wang2024b}, the precursor flare of AT2021aeuk is separated from the first major flare with a longer duration of $\sim$200 days. 
One hypothesis is that this precursor event is entirely independent of the transient and triggered by random activity of the AGN. However, if it is related to the dual-flare activity, simulations of TDEs within AGNs suggest a mechanism \citep{Chan2019}: if the TDE fallback stream is dense enough to penetrate the pre-existing AGN disk, a small precursor flare can be triggered due to minimal kinetic energy loss during this penetration. Most of the fallback stream's energy is preserved and subsequently released during a second interaction between the TDE stream and the AGN disk, resulting in the first major flare in our target. However, the absence of a precursor flare before the second flare suggests that the TDE stream may not penetrate the AGN disk during the second disruption, releasing all kinetic energy in a single interaction. This could be due to a smaller amount of stripped gas or a different trajectory, resulting in a different interaction radius during the second disruption.

\begin{figure}
    \begin{center}
        \includegraphics[width=0.5\textwidth]{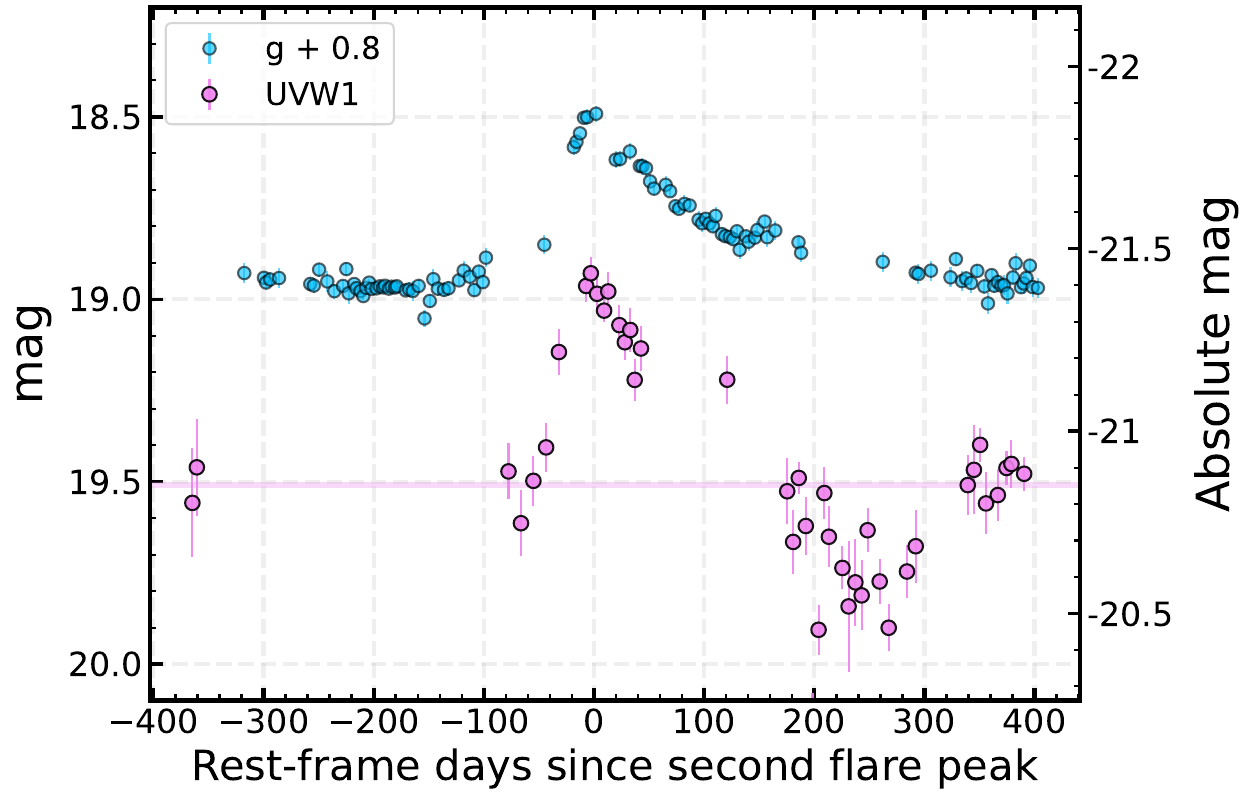}
        \caption{The $UVW1$ light curve for the second flare is presented without host subtraction. The pre-outburst brightness, estimated through SED fitting, is denoted by the horizontal line, which aligns with the observed $UVW1$ luminosity at both the end of the first flare and the onset of the second. The $UVW1$ light curve became fainter than the pre-outburst level at the end (200 days) of the second flare, and recovered to previous baseline within 200 days. In contrast, the $g$-band light curve exhibits a consistent, smooth decline.
        \label{fig:second_flare}}
    \end{center}
\end{figure}

The third remarkable feature is the UV dip. Figure~\ref{fig:second_flare} shows the $UVW1$ light curve up to 400 days after the second flare. The photometry at the end of the first flare is considered as the baseline for $UVW1$ in the quiescent state. This is consistent with the pre-outburst UV brightness estimated from the SED fitting, suggesting a reliable estimate of the quiescent flux level. Notably, around 200 days after the peak of the second flare, the $UVW1$ light curve drops below the pre-outburst level, continuing to decline to its faintest state at around 250 days, approximately 0.3 magnitudes fainter than the previous brightness. The $UVW1$ light curve then recovers to the pre-outburst level within the next 100 days. Similar behavior has been observed in the optical light curves of other transients, such as CSS100217 and AT2019brs. However, the V-band light curve of CSS100217 remained 0.4 magnitudes fainter than the pre-outburst level and did not recover, even in observations taken 1000 days after the flare peak in 2013 and no further data \citep{Cannizzaro2022}. In the case of AT2019brs \citep{Frederick2021}, the decline continued, reaching over one magnitude below the pre-outburst level by February 2024, according to the ZTF $g$-band light curve. This type of dip following a flare may be interpreted as the destruction of the accretion disk by the fallback stream within the TDE framework \citep{Cannizzaro2022}, while the optical dip may suggest a more extensive deficit region that requires a longer time to recover.

\subsubsection {Time Lag Detection}\label{sec:lag}
\begin{figure}
\begin{center}
    \includegraphics[width=0.5\textwidth]{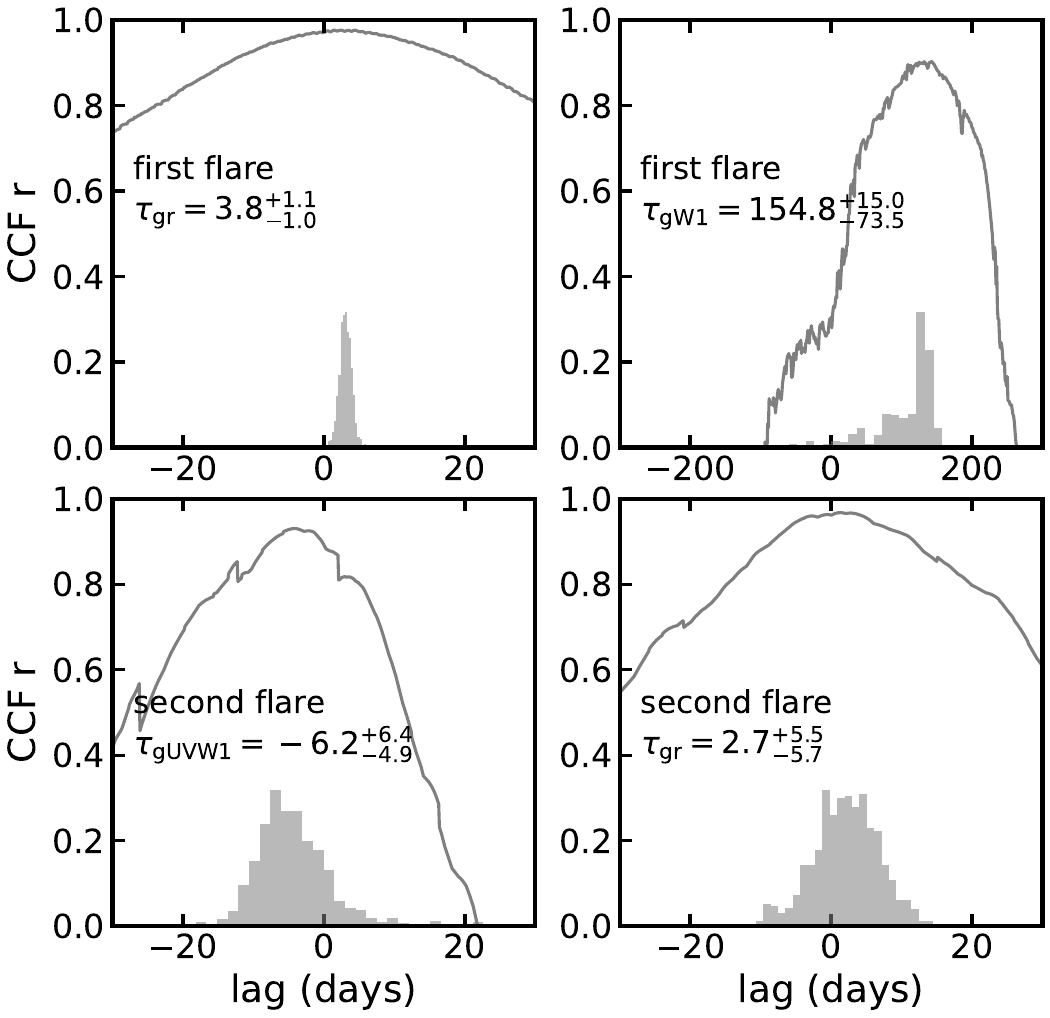}
    \caption{The ICCF results for the first and second flares between the multi-band light curves. The posteriors of the centroid lags are shown as gray histograms, while the black lines represent the CCF curves. The measured lags and their uncertainties are listed in each panel, in units of days. }
    \label{fig:iccf}
\end{center}

\end{figure}

We employed the interpolated cross-correlation function (ICCF) method \citep{Gaskell1987, White1994} to calculate the continuum time lags across UV/optical/MIR light curves for two distinct flares.  The code PyCCF \citep{Sun2018} is used for the ICCF calculations. The search range is $-$30 to 30 days for UV/optical and $-$300 to 300 days for MIR. Uncertainty estimation was performed using the standard flux randomization/random subset selection (FR/RSS) method described in \cite{Peterson1998}. The $g$-band light curve serves as the reference given its relatively stronger variability, coupled with excellent sampling and high photometric accuracy. The results are presented in Figure~\ref{fig:iccf}. In both flares, the continuum time lag increases with wavelength, aligning with continuum reverberation mapping (RM) results in both AGNs \citep{Fausnaugh2016,Cackett2018,Guo2022} and TDEs \citep{Guo2025,Faris2024}. The MIR lag is $\sim$155 days in the flares, consistent with the prediction by the $R_{\rm torus}-L_{\rm 5100}$ relation in AGN \citep{Chen2023}, given $L_{\rm 5100} = 1.5\times 10^{44}\rm \ erg\ s^{-1}$. The MIR lag for the second flare is not well constrained due to the sparse MIR light curve, which fails to capture the peak.

\begin{figure}[!]
    \centering
    \hspace{-1cm}
    \includegraphics[width=0.5\textwidth]{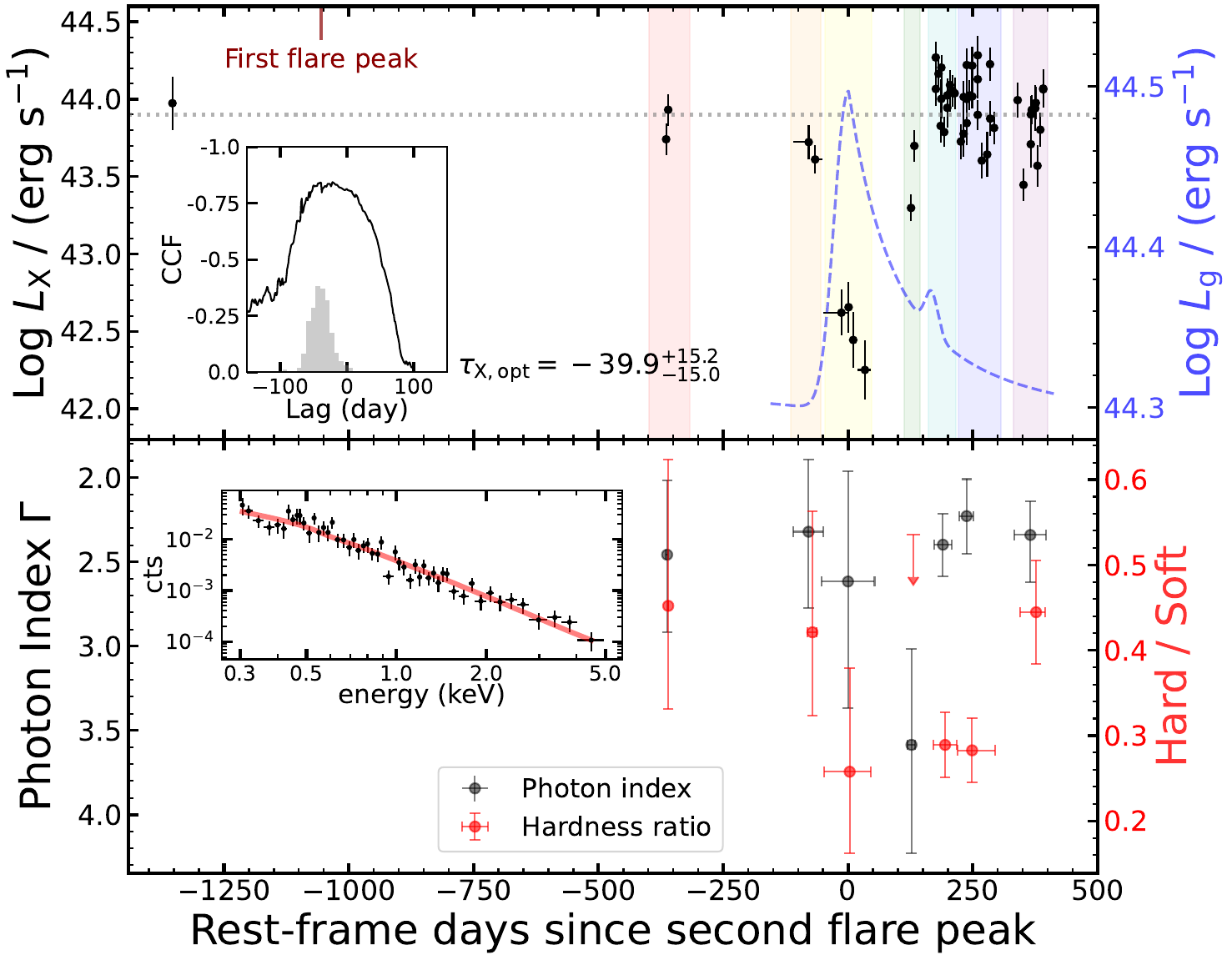}
    \caption{\textit{Top panel:} Evolution of the X-ray light curve during the second flare, with different bins color-coded. The inset panel displays the measured time lag demonstrating the anti-correlation between the optical and X-ray light curves. \textit{Bottom panel:} Evolution of the X-ray photon index (black) and the hardness ratio (red). Despite significant error bars, a trend of ``harder-when-brighter" is clued. The inset panel shows the composite X-ray spectrum, modeled by a power-law with a reduced $\chi_{v}^{2}$ = 1.2.}
    \label{fig:index}
\end{figure}

\subsection{X-ray evolution}\label{sec:x-ray}     

One remarkable characteristic of AT2021aeuk is the anti-correlation between X-ray and UV/optical variability during the second flare, as shown in Figure \ref{fig:index}. Swift/XRT multi-epoch observations before and after the first flare indicate a constant luminosity around $10^{43.9}\rm \ erg\ s^{-1}$. Following the onset of the second flare, the X-ray luminosity began to decrease sharply, displaying a V-shaped variability pattern and resulting in a strong anti-correlation with an ICCF coefficient of $r_{\rm max} \sim -0.8$ relative to the optical flare. The reversed optical flare leads the X-ray dip by $39.9^{+15.2}_{-15.0}$ days. Within 150 days, the X-ray emission recovers to a luminosity comparable to its pre-outburst state. This X-ray behavior is distinct from that typically seen in AGNs or TDEs within quiescent galaxies, yet it resembles patterns observed in several TDE candidates in AGNs. Further details will be discussed in section \ref{sec:diss_x}.




To investigate the evolution of the X-ray spectral shape, we grouped the X-ray observations into six bins, balancing photon counts and representative periods: the quiescent stage after the first flare, the onset of the second flare, the faintest stage, the rapid recovery, and two post-flare stages. The time ranges for these bins were manually selected to optimize the spectral SNR, minimize flux variation, and shorten the duration of each epoch. Each spectrum was effectively fitted with a power-law model plus foreground Galactic absorption, suggesting a non-thermal origin likely linked to coronal emission. The power-law indices and the ratio of hard X-ray (1.5$-$10 keV) to soft X-ray (0.3$-$1.5 keV) for each period are illustrated in Figure~\ref{fig:index}. The evolution of the spectral index and the hard-to-soft ratio both exhibit a V-shaped pattern, similar to the X-ray light curve variation, demonstrating a potential ``harder-when-brighter" trend, given the relatively large uncertainties. 


\begin{figure*}
\begin{center}
    \includegraphics[width=0.9\textwidth]{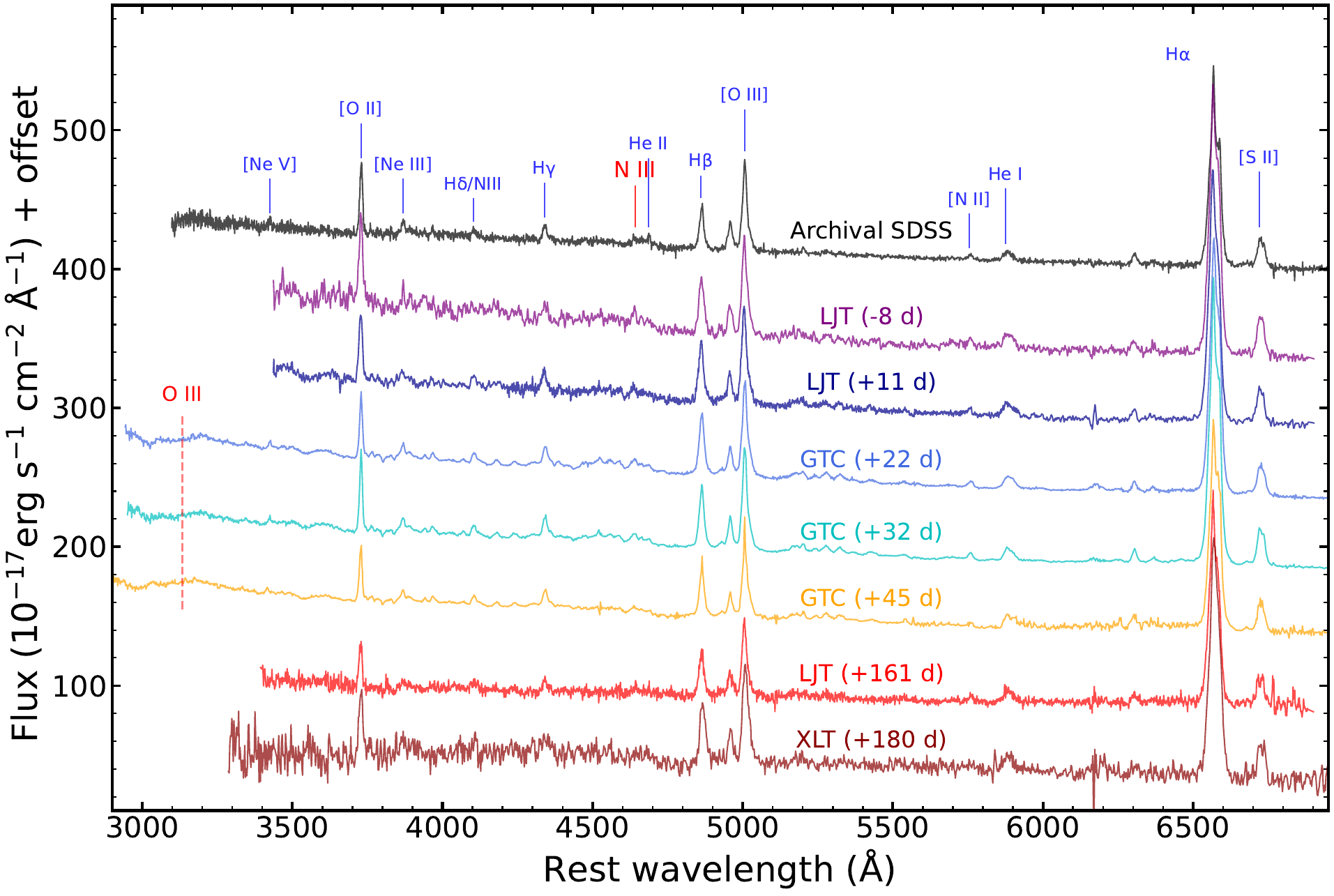}
    \caption{Spectroscopic follow-up observations of AT2021aeuk during its second flare. All spectra have been corrected for Galactic extinction. For comparison, the SDSS archival spectrum from 2003 is included as a reference. The positions of the fluorescence lines \ion{O}{3}$\lambda 3133$ and \ion{N}{3}$\lambda 4641$ are highlighted with red dashed and solid lines, respectively. The observation times are indicated relative to the second flare peak in the rest frame.}
    \label{fig:spectra}
\end{center}
\end{figure*}

\begin{figure}
    \includegraphics[width=0.45\textwidth]{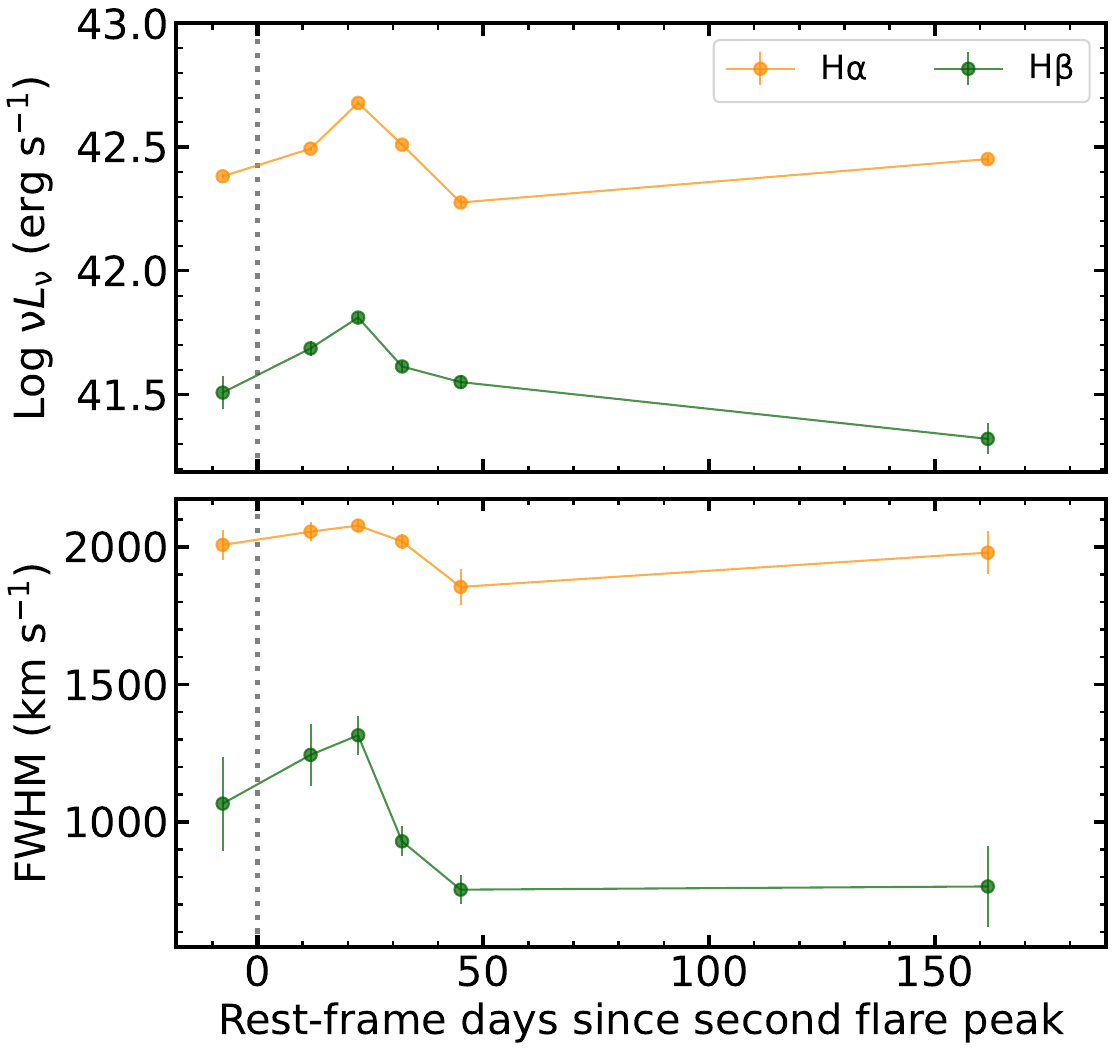}
    \caption{The luminosity and FWHM evolution of the $\rm H\alpha$ and $\rm H\beta$ lines. Both parameters display a delay of $~$25 days to the second flare. Note that the XLT epoch is discarded due to significant influence
     from moonlight, poor seeing conditions, and high airmass.}
    \label{fig:line_lc}
\end{figure}

\begin{figure}
\begin{center}
    \includegraphics[width=0.45\textwidth]{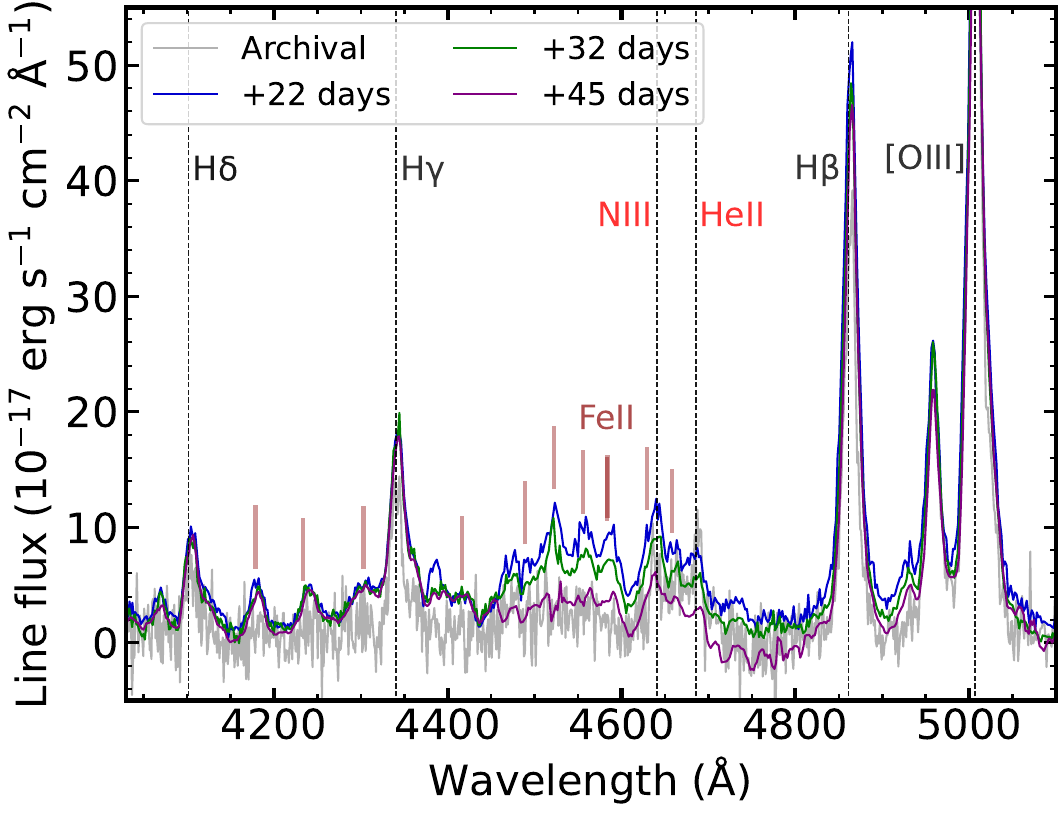}
    \caption{The continuum-subtracted spectra from archival SDSS (gray line) and GTC observations (blue/green/purple lines). The short red lines indicate \ion{Fe}{2} emission lines. The variability of \ion{N}{3} and \ion{He}{2} lines is not discernible after subtracting the \ion{Fe}{2} pseudo continuum. }
    \label{fig:bowen}
\end{center}
\end{figure}

\subsection{Spectroscopic Characteristics}\label{sec:sepc}
All calibrated spectra are presented in Figure~\ref{fig:spectra}. During the second flare, the overall continuum slopes show minimal variations, as they are already dominated by the AGN's blue continuum. As depicted in Figures~\ref{fig:h_alpha} and \ref{fig:h_beta}, new broad Balmer lines emerge with relatively symmetric and virialized profiles. These do not resemble typical TDE features, which are characterized by extremely broad line widths ($\mathrm{FWHM > 10^4\ km\ s^{-1}}$) and significant variability in flux, profile, and velocity offset \citep[e.g.,][]{Charalampopoulos2022}. The luminosity evolution of the broad H$\alpha$ and H$\beta$ lines in Figure~\ref{fig:line_lc} exhibits a lag of $\sim$ 25 days relative to the optical peak, aligning with RM results in both AGNs and TDEs \citep{Peterson1998,Charalampopoulos2022}. Moreover, the observed lag is consistent with the $R_{\mathrm{BLR}}-L_{5100}$ relation for AGNs with high accretion rates \citep{Du2018}, given a continuum luminosity of $L_{5100} = 1.4 \times 10^{44}$ erg s$^{-1}$. This probably suggests that the new-born broad component is associated with pre-existing broad-line clouds. However, the correlated luminosity and line width evoke a ``broader-when-brighter" evolution typically observed in TDEs \citep{Roth2016,Charalampopoulos2022}.

On the other hand, both the \ion{He}{2} $\lambda 4686$ emission line and the Bowen fluorescence line \ion{N}{3} $\lambda 4641$ are detected in the SDSS archival spectrum, indicative of exposure to extreme UV radiation \citep{Bowen1928}. However, subtle variations are evident (see Figure~\ref{fig:bowen}), which differ from those typically observed in Bowen fluorescence flares \citep[e.g.,][]{Tadhunter2017, Trakhtenbrot2019a, Makrygianni2023,Veres2024}. Additionally, the \ion{Fe}{2} emission lines appear to respond to variations in the continuum, suggesting the release of the Iron element from the inner region of the dust torus \citep{He2021}.

The absence of extremely broad emission lines distinguishes AT2021aeuk from typical TDEs in quiescent galaxies. However, the diverse range of TDE spectra, including a featureless population, highlights the complexity of understanding emission line behaviors in these events \citep[e.g.,][]{Roth2016, Lu2020, Hammerstein2023b, Yao2023}. Notably, very broad lines are rare in TDEs associated with AGNs, with only one exception of F01004-2237, where the AGN's pre-outburst activity was minimal or had ceased \citep{Tadhunter2017, Sun2024}. We therefore speculate that TDEs in AGNs might inherently struggle to produce very broad emission lines, possibly due to the efficient re-accretion of disrupted debris onto the disk, as the failed outflow in AGN \citep{Czerny2011}.

\section{The nature of AT2021aeuk}\label{diss:flare}

\subsection{Explanation for the X-ray Evolution}
\label{sec:diss_x}

The drastic X-ray evolution during the second flare of AT2021aeuk is remarkable for its anti-correlation with the optical flare, accompanied by a 40-day delay, and is dominated by a power-law component throughout the entire evolution. This behavior contrasts with the typical X-ray characteristics seen in AGNs or TDEs. In AGNs, X-ray and optical light curves generally show a positive correlation, with X-ray variability leading the optical by several days at most. For TDEs, X-ray spectra are mostly dominated by a thermal component if detected, where the hard X-ray corona could form in the late time \citep[e.g.,][]{Guolo2024}. While the correlation between X-ray and optical in TDEs is generally unclear, based on a handful of well-monitored TDEs, X-ray emission usually lags behind the optical flare by tens to hundreds of days, likely due to the delayed accretion disk formation \citep[e.g.,][]{Guo2025} or the declining obscuration \citep[e.g.,][]{Dai2018}.


\begin{figure}
\hspace{0cm}
    \centering
    \includegraphics[width=0.5\textwidth]{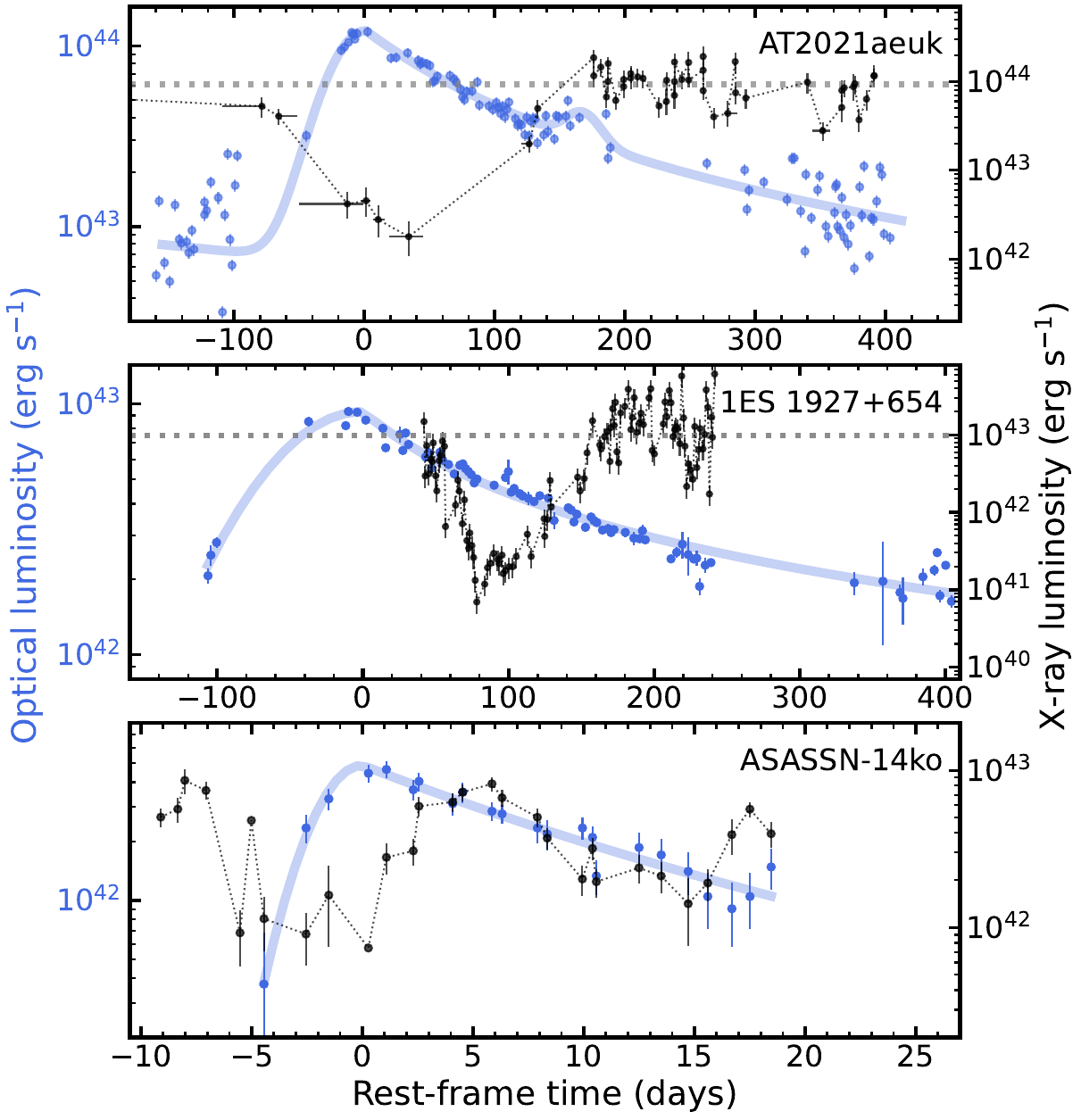}
    \caption{The X-ray dip (black) and the optical flare (blue) for three TDE candidates in AGNs. The horizontal dotted line represents the quiescent-state X-ray luminosity, which is absent in the bottom panel because ASASSN-14ko has been exhibiting repeating flares since discovery and it is hard to determine a quiescent state for the X-ray.}
    \label{fig:compare_x-ray}
\end{figure}

Interestingly, aside from AT2021aeuk, a V-shaped X-ray evolution accompanied by an optical flare has also been observed at least in two other TDE candidates in AGNs: 1ES 1927+654 \citep{Trakhtenbrot2019b} and ASASSN-14ko \citep{Payne2021}, as shown in Figure~\ref{fig:compare_x-ray}. The pre-outburst X-ray luminosity of 1ES 1927+654, characterized by both a power-law and a blackbody component, is interpreted as originating from the corona and the innermost regions of the accretion disk, respectively \citep{Boller2003,Gallo2013,Cao2023}. Approximately 50 days after the optical peak, the power-law component disappeared, and the X-ray luminosity continued to decline, with the spectral hardness decreasing. The X-ray reached its faintest level about 80 days after the flare peak, with a luminosity roughly 100 times fainter than its pre-outburst state. Over the next 100 days, the X-ray luminosity returned to the pre-outburst state, accompanied by the gradual recovery of the power-law component in the X-ray spectrum \citep{Ricci2020}. In another case, the ASASSN-14ko exhibits repeating UV/optical flares with a cycle of $\sim$ 115 days, with each flare evolving rapidly over $\sim$30 days \citep{Payne2021}. The X-ray luminosity of the source dropped to its faintest level almost concurrently with the peak of the optical flare \citep{Payne2022,Huang2023b,Payne2023}. Simultaneously, the X-ray spectra hardened as the emission declined, without any variations in absorption \citep{Payne2023}. This ``harder-when-fainter" pattern resembles the typical X-ray variability in X-ray binary and AGN, while differs from the behavior seen in AT2021aeuk and 1ES 1927+654, possibly indicative of a different physical origin.

The optically thick obscuration is unlikely to account for the X-ray evolution, as this scenario would predict that the inverted X-ray light curve should coincide with the optical flare. This contradicts the observed lags of several tens of days between the X-ray and the inverse optical signals, particularly in the cases of AT2021aeuk and 1ES 1927+654. Furthermore, there is a lack of supporting evidence from X-ray absorption caused by obscuration. Alternatively, \citet{Scepi2021} suggested that an influx of accreting material carrying inverse magnetic flux could cut off the energy supply to the corona, if the corona is powered by the Blandford-Znajek process. However, for example in AT2021aeuk, with a BH mass of $10^{6.9}\ M_\odot$ and an Eddington ratio of 0.5, the UV/optical emitting region at a temperature of $10^{4}\ \rm K$ is located around 2000$R_{\rm g}$. Within the observed lag of $\sim$40 days between the optical and X-ray emissions, it is unlikely that the optical emitting region in an accretion disk could directly influence the X-ray corona, assuming perturbations travel at sound speed. This issue becomes even more severe for ASASSN-14ko, given the near-zero lag.




\begin{figure*}[htbp!]
    \centering
    \includegraphics[width=0.9\textwidth]{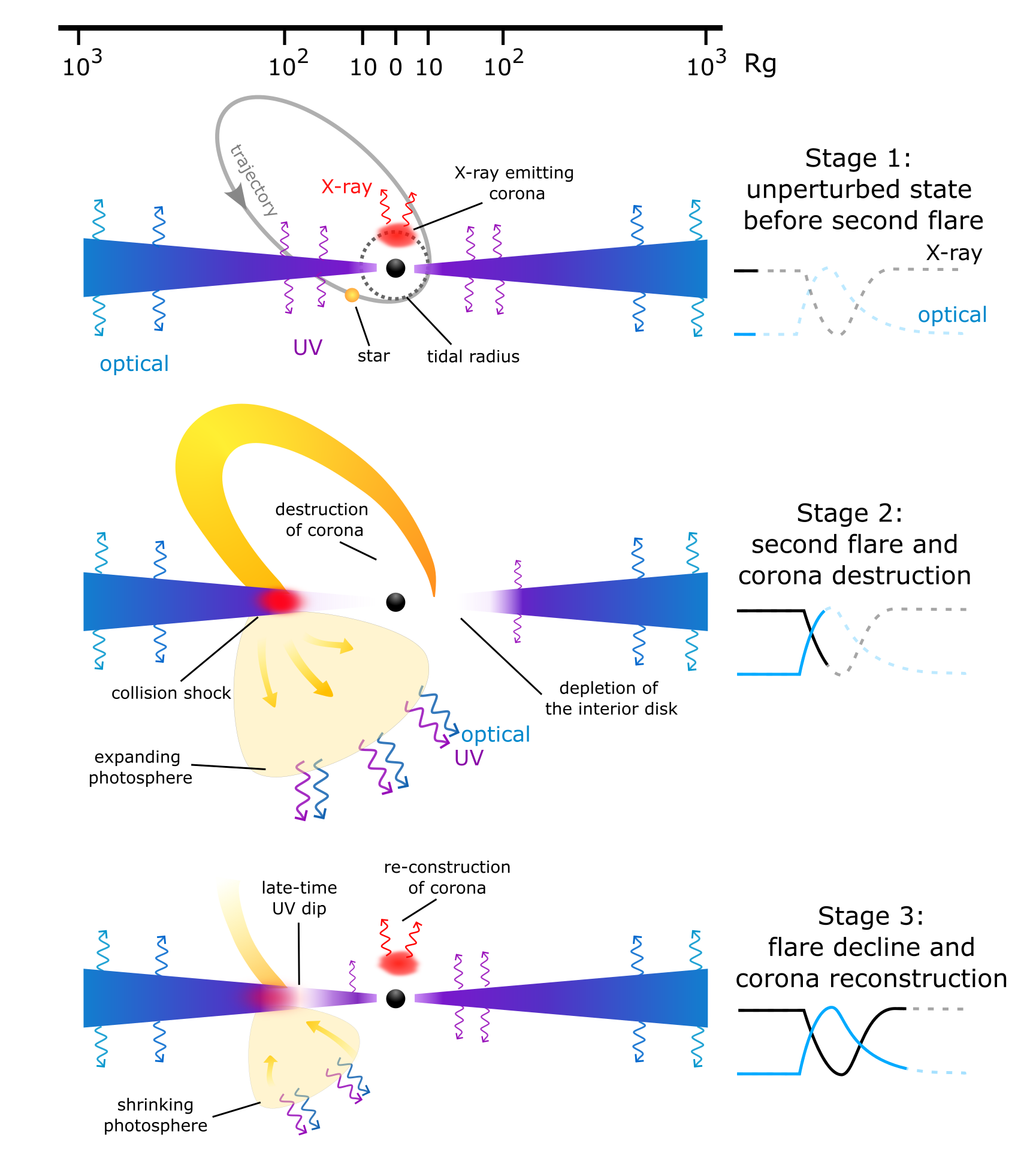}
    \caption{A brief scheme about the stream-disk evolution to explain the anti-correlation between the optical and X-ray light curves in the second flare. \textit{Top panel:} The trajectory of the disrupted star intersects with the inner disk. \textit{Middle panel:} Following an elliptical orbit, the fallback stream hits the disk and releases energy through the shock, heating the expanding photosphere, which produces immense UV/optical emission. The strong advection induced by the stream-disk collision depletes the interior disk, and destroys the X-ray corona, causing the X-ray decline. \textit{Bottom panel:} In the late time, the decreasing fallback rate leads to the decline of the UV/optical flare. The corona is rapidly re-constructed, followed by the recovery of the accretion disk region closer to the collision point. }
    \label{fig:cartoon}
\end{figure*}


Thus, what mechanism drives the observed optical and X-ray behaviors? One possible scenario is that the X-ray corona is (partially) destroyed by the TDE fallback stream, which collides with the inner part of the accretion disk, close to the corona. The nearby corona could either disappear due to the depletion of the inner accretion disk that disrupts the material supply \citep{Uttley2014}, or it might be directly blown away by strong winds caused by the increased accretion rate \citep{Cao2023}. According to the simulation \citep{Chan2019}, the shock generated by the stream-disk collision can dissipate kinetic energy into UV/optical emission with near-Eddington bolometric luminosity, supporting the disk depletion scenario.

Figure~\ref{fig:cartoon} illustrates a potential stream-disk collision scenario for AT2021aeuk. Involving a BH with a mass of $10^{6.9}\ M_{\odot}$, the trajectory of the disrupted star intersects with the inner disk (Stage 1). Once disrupted, the elongated debris stream follows an eccentric orbit before falling back to the BH. The collision between the fallback stream and the accretion disk triggers the observed UV/optical flare and leads to the destruction of the X-ray corona and the inner accretion disk (Stage 2). As the fallback rate declines, the corona begins to reform, followed by the restoration of the disk region near the collision point (Stage 3).

We roughly estimate the distance of the collision point from the BH based on a pre-existing thin disk \citep{Shakura1973}. We assume the pericenter $R_{\rm p}$ equal to the tidal disruption radius $R_{\rm t}$, as: $R_{\rm p} \simeq R_{\rm t} = R_{*}\left( M_{\rm BH}/M_{*} \right)^{1/3}$, 
where the $M_{*}$ and $R_{*}$ are the stellar mass and radius. In the assumption of a solar-type star with $M_{*}=M_{\odot}$ and $R_{*} = R_{\odot}$, the pericenter is around 12$R_{\rm g}$ for a $10^{6.9} M_{\odot}$ BH. Then the eccentricity $e_{\rm min}$ of the innermost orbit is given by \cite{Bonnerot2021}:
\begin{equation}
    1 - e_{\rm min} = 0.02 \left(\frac{M_{\mathrm{BH}}}{10^{6} \mathrm{M}_{\odot}}\right)^{-1 / 3} \approx 0.01
    \end{equation}
The distance of the two projective intersection points can be calculated as:
\begin{equation}
R_{c} = \frac{a^{2}-f^{2}}{a \pm f\sin \theta}
\end{equation}
where $R_{\rm c}$ is the distance between the collision point from the BH, $a$ is the semi-major axis of the elliptical orbit, equal to $R_{\rm p}/(1-e_{\rm min})$, $f$ is the focal distance of the elliptical trajectory, equal to $a-R_{\rm p}$, $\theta$ is the inclination angle between the disk pole and the stream orbit. Assuming an inclination of 60 degrees, which is the median inclination for randomly orienting stars, the distance of the closer and farther collision points is around 1 and 14$R_{\rm t}$, corresponding to 12 and 160$R_{\rm g}$, respectively. Therefore, the collision point is likely located at the inner disk, generally consistent with the observed short lags between the optical and X-ray emission.


Furthermore, simulation results suggest that the accretion disk interior to the collision point may become depleted due to rapid inflow caused by kinetic dissipation \citep{Chan2019}. Assuming the collision point is around 70$R_{\rm g}$, this corresponds to a disk temperature of $\sim 10^{5}$ K or UV emission at $\sim$1400 \AA, which means the stream collision in the UV emitting region of the disk could potentially explain the observed late-time dip in the $UVW1$ light curve. In contrast, the optical emission, primarily originating from the outer accretion disk, is less affected, resulting in the optical flare gradually returning to the quiescent level.

\begin{deluxetable*}{lcccccccccc}
\tablecaption{{Seven optical rpTDE candidates} } 
\tablehead{\colhead{Name} & \colhead{$z$} & \colhead{Host type} & \colhead{log $M_{\rm BH}/M_{\odot}$} & \colhead{$N_{\rm flare}$} & \colhead{Period/yr} & \colhead{$E_{\rm flare2}$/$E_{\rm flare1}$} & \colhead{$t_{\rm third}$} & \colhead{Reference}}
\startdata
ASASSN-14ko & 0.0425 & Type2 AGN & 7.9 & $>21$ & 0.31 & $\sim 1$ & $-$ & \cite{Payne2021,Payne2022,Payne2023} \\
AT2018fyk & 0.0590 & quiescent & 7.7 & 2 & 3.3 & 0.1 & 2025/03 & \cite{Wevers2023,Wen2024} \\
AT2020vdq & 0.0450 & quiescent & 6.1 & 2 & 2.5 & 30 & 2026/01 & \cite{Yao2023,Somalwar2023a} \\
AT2022dbl & 0.0284 & quiescent & 6.4 & 2 & 1.9 & 0.7 & 2026/01 & \cite{Lin2024} \\
AT2019aalc & 0.0356 & Sey1 & 7.2 & 2 & 3.9 & $>1.3$ & 2027/08 & \cite{Veres2024} \\
F01004-2237 & 0.1178 & AGN & 7.4 & 2 & 10.3 & 0.05-0.5 & 2033 & \cite{Tadhunter2017,Sun2024} \\
AT2021aeuk & 0.2336 & NLS1 & 6.9 & 2 & 2.9 & 0.4 & 2026/09 & This work 
\enddata
\tablecomments{$N_{\rm flare}$: the number of exhibited flares to date. $E_{\rm flare2}/E_{\rm flare1}$: the ratio of total released energy between the latter and the first flare. For AT2021aeuk and AT2019aalc, we adopt $g$-band monochromatic energy as no UV observations are available in their first flares. $t_{\rm third}$: the predicted time of the third flare, assuming a same period between the first and second flares.}
\label{tab:rpTDE}
\end{deluxetable*}

\subsection{The Repetition of Two Similar Flares}


The similar light curve shapes and color evolution of the two flares observed in AT2021aeuk suggest they may be related. Given the TDE rate of $\sim 3.2\times10^{-5} \rm \ yr^{-1}\ galaxy^{-1}$ \citep{Yao2023}, the probability of observing an additional TDEs within 3 years is around $10^{-4}$. This probability drops dramatically to 10$^{-8}$ if two independent flares within such a time baseline are searched for directly in the light curve database. Despite factors such as starburst activity \citep{Stone2016,Stone2018}, the accretion disk \citep{Kaur2024}, or the presence of SMBH binary \citep{Ivanov2005} might enhance the TDE rate, the entire flare rate in AGNs remains low ($<$ 0.1\%), as evidenced by studies of flaring AGNs \citep{Graham2017} and extreme variability quasars \citep{Ren2022}. Therefore, it is unlikely that the observed outburst in AT2021aeuk is related to two independent flares, although this cannot be entirely excluded.

\begin{figure}
    \includegraphics[width=0.45\textwidth]{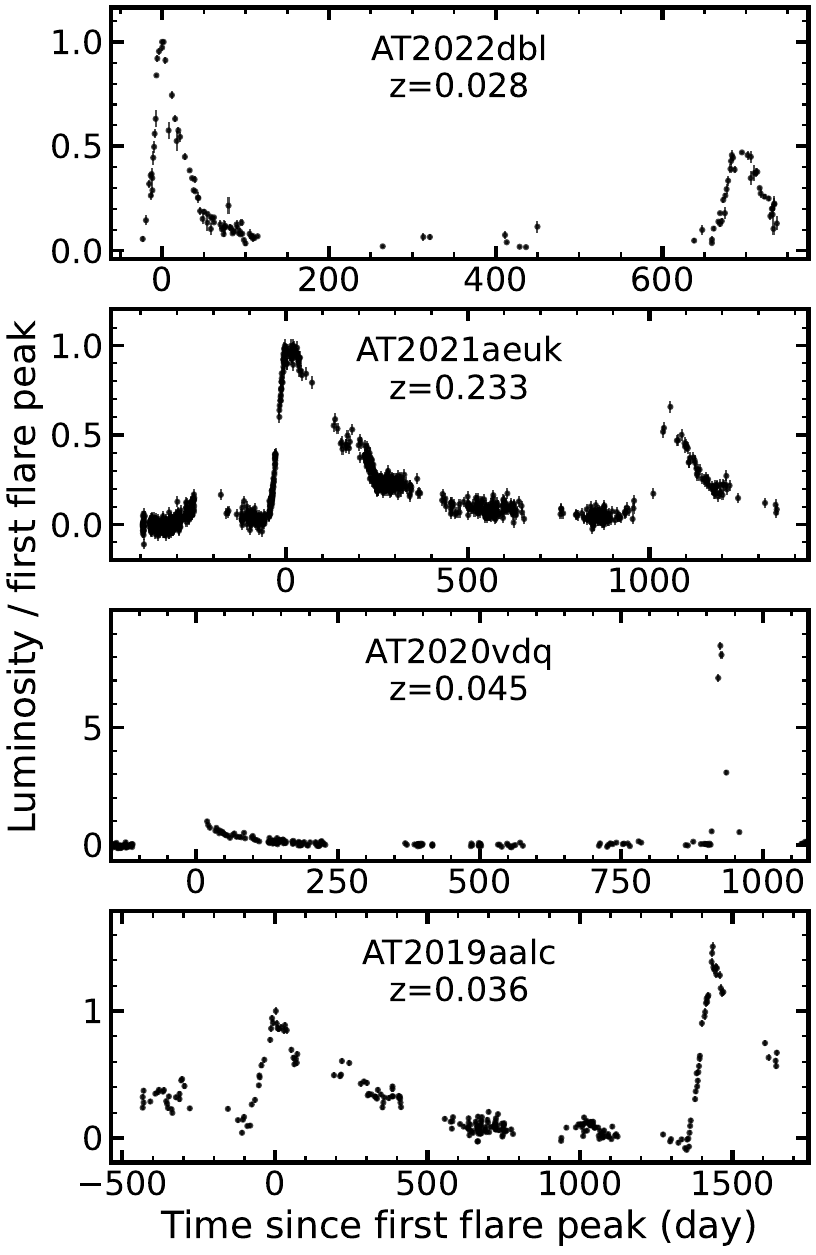}
    \caption{The four rpTDEs within the ZTF footprint. The AT2020vdq and AT2022dbl are rpTDEs in quiescent galaxies, while the AT2019aalc and AT2021aeuk are rpTDE candidates in AGN. The y-axis is set as the ratio between evolving luminosity and the peak luminosity of the first flare.}
    \label{fig:rpTDE}
\end{figure}

\subsubsection{Repeating Flares in TDE Scenario} 
If the second flare of AT2021aeuk originates from a TDE, repeating flares could potentially be explained by a rpTDE, supported by the similar light curve shapes and color evolution observed in both flares. To date, 7 optical \citep{Payne2021,Payne2022,Payne2023,Huang2023b,Wevers2023,Somalwar2023a,Lin2024,Veres2024,Sun2024} and 3 X-ray \citep{Miniutti2023,Malyali2023,Liu2024_XrpTDE} recurring flares are identified as rpTDE candidates. Among them, AT2022dbl stands out as the most reliable candidate given its similarities in light curve evolution and blackbody temperature, particularly the identical spectral emission lines during two flares \citep{Lin2024}. In this work, our discussion focuses on the optical rpTDE candidates, with detailed information presented in Table \ref{tab:rpTDE}.

ASASSN-14ko stands out among rpTDE candidates due to its rapid evolution time of approximately 30 days, low energy release of $E\sim 3 \times 10^{50}$ ergs, and a short recurrence period of about 115 days between flares. This pattern suggests that a star may periodically graze the tidal boundary, stripping only a small portion of its envelope to feed the BH with each pass \citep[e.g.,][]{Cufari2022, Bandopadhyay2024, Liu2024}. As a result, the orbit and structure of the star vary little in each tidal encounter, yielding numerous similar repetitions. 


Among the six other rpTDE candidates, variations in the relative amplitude of the second flare compared to the first are different. AT2018fyk, AT2022dbl, F01004-2237, and AT2021aeuk exhibit a weaker second flare, maintaining a similar light curve shape to the first. In contrast, AT2020vdq and AT2019aalc show a more luminous second flare with a sharper rise and fall. Figure~\ref{fig:rpTDE} illustrates the light curves of four rpTDE candidates within the ZTF footprint, given the uniform data quality. The variation in the second flare's behavior indicates a different striped mass in each tidal encounter, possibly associated with the change in the orbit \citep{Chen2024} and stellar condition \citep{Ryu2020}. An attenuating repeating flare pattern is consistent with the remnant gaining orbital energy, achieving a larger pericenter after each disruption \citep{Manukian2013, Gafton2015}, and ultimately settling into a safe trajectory after a series of progressively weaker flares. In contrast, more mass could be stripped out in the next tidal encounter if a star is losing orbital energy \citep{Ryu2020, Chen2024}, or becoming more vulnerable due to its spin-up \citep{Bandopadhyay2024} and structural changes \citep{Liu2024}, leading to a larger tidal radius after each disruption. In this scenario, each tidal encounter could accelerate the star towards complete disruption. Therefore, the observations of a third flare from these rpTDE candidates could verify distinct scenarios and provide constraints on the dynamic models. The predicted third flare of AT2021aeuk is expected in September 2026, assuming a similar period.

\subsubsection{Stellar-mass Binary Black Bole Merger within an AGN Disk}

Another possible interpretation of the repeating flares is the electromagnetic counterpart of a stellar-mass binary black hole (BBH) merger within an AGN disk \citep[e.g.,][]{Graham2020_sBH}. The gravitational wave radiation from merging unequal mass BBHs leads to a recoil of the BBH center of mass. The kicked bounded sphere of the merger product will collide with the surrounding AGN disk, accounting for the precursor flare. Followed by the kicked merging product, the unperturbed gas surrounding the stellar-mass black hole (sBH) is accreted, producing the first UV/optical flare through the shock reprocessing of the accretion material. Finally, the kicked BH will fall back to the disk as the kick velocity is smaller than the local escape velocity, resulting in the second flare through the BH-disk collision. 

We perform some quantitative estimations of the luminosity and timescales for the series of flares.
\begin{enumerate}
    \item \textbf{precursor flare from the kicked material:} For a kicked velocity $v_{\rm k}$ of the merged BBH with a mass of $M_{\rm BBH}$, the bounded gas around the BBH is $M_{\rm b}\simeq 4\pi/3\rho R_{\rm b}^{3}$ with $R_{\rm b}={\rm min}[GM_{\rm BBH}/v_{\rm k}^{2},\ (M_{\rm BBH}/3M_{\rm smbh})^{1/3}r]$. The total energy dissipation during the collision $E_{\rm b}=1/2M_{\rm b}v_{\rm k}^{2}$ released in a dynamical timescale of $t_{\rm ram}=R_{\rm b}/v_{\rm k}$. With $v_{\rm k}=100\ {\rm km\ s^{-1}}$, and $M_{\rm BBH}=100\ M_{\odot}$, the luminosity of the shocked material during the merger kick is $\sim10^{43}\ {\rm erg\ s^{-1}}$, and the duration is $\sim 150$ days, roughly consistent with the precursor flare of AT2021aeuk.
    
    \item  \textbf{First flare from the sBH accretion:} We approximate the accretion of the sBH as a Bondi-Hoyle-Lyttleton (BHL) process. The total accretion luminosity is $L_{\rm BHL}=\eta \dot{M}_{\rm BH} c^2$,  where $\eta=0.1$ is the radiative efficiency, $\dot{M}_{\rm BHL}$ is the BHL accretion rate. We neglect the mass loss during the merging process such that $M_{\rm BH}\simeq M_{\rm BBH}$. As the BH enters into an unperturbed disk region, which could be far away from the middle plane, we assume a gas density of $\rho=10^{-10}\ {\rm g\ cm^{-3}}$.  With the parameter adopted above, we can estimate that $L_{\rm BHL}\sim 10^{45}\ {\rm erg\ s^{-1}}$. 
    The first flare will end at the timescale of $t\sim {\rm max} [t_{\rm diff}, H/(v_{\rm k}\sin\theta)]\simeq 500\ {\rm days}$, where $t_{\rm diff}\simeq H\tau/c$ is the photon diffusion timescale with $\tau\sim10^{4}$ being the optical depth of the accretion flow, $\theta=60^{\circ}$ is the kicked angle relative to the disk midplane, and $H/r\simeq 0.01$ at $r\simeq 10^{4}{R_{\rm g}}$ from the SMBH.
    
    \item \textbf{Second flare from BH-disk collision:} Assuming the merged product is kicked out of the disk at $\sim 10^{4}R_{\rm g}$, this sBH will fall back to hit the disk after a time interval of the local orbital period of $\sim P_{\rm orb}/2\simeq 1400\ \rm days$, broadly consistent with the separation between the precursor flare and second flare. The shocked luminosity during the BH-disk collision is on the order of $L_{\rm col}\sim 10^{44}\ \rm erg\ s^{-1}$ in a similar duration as the first flare. 
\end{enumerate}
In this scenario, the decline in X-ray emission during the second flare could be attributed to the outflow induced by the BH-disk collision, which may blow away the corona with its strong outflow. If the second flare occurs at a distance of $\sim 10^{4}R_{\rm g}$, the observed time lag of 40 days between the optical and X-ray requires an outflow speed of approximately 0.34 $c$. Combining the effects of both shock and accretion, the BH-disk collision is capable of launching a much stronger outflow compared with the stream-disk collision. One prediction of this scenario is that the merged BH will periodically impact the disk, producing repeating flares with the period and amplitude modulated by the BH-disk interaction \citep{Gilbaum2024}.

\section{Conclusions}
\label{sec:conclusion}
We report the occasional discovery of a repeating flare event, AT2021aeuk, in an NLS1 galaxy at $z = 0.2336$. To investigate its underlying physics, we analyzed multi-wavelength spectroscopic and photometric observations spanning from X-ray to MIR bands during the second flare. The key findings are summarized as follows:
\begin{enumerate}
    \item AT2021aeuk exhibits two prominent, similar, and relatively smooth flares within $\sim$3 years, accompanied by a precursor flare, in a previously stable NLS1 galaxy (Figure \ref{fig:long_lc}). The luminosity ratios for the first, second, and precursor flares are around $1:0.65:0.2$, respectively.

    \item The multi-band photometric monitoring during the second flare reveals a remarkable anti-correlation between the X-ray and UV/optical light curves, with the X-ray lagging behind the optical by 39.9$^{+15.2}_{-15.0}$ days (Figure \ref{fig:lc} and \ref{fig:index}). This observation is inconsistent with pure AGN flare or SN scenarios as the origin of the second flare.
    
    \item The constant color index (Figure \ref{fig:compare_lc}) or blackbody temperature (Figure \ref{fig:TBB}), and positive UV/optical/MIR lag direction (Figure \ref{fig:iccf}) suggest that both flares resemble TDEs. 
    
    \item New-born broad components have been observed during the second flare, with FWHMs of $\sim$1000 km s$^{-1}$ for H$\beta$ and 2000 km s$^{-1}$ for H$\alpha$ (Figures \ref{fig:spectra}, \ref{fig:h_alpha}, and \ref{fig:h_beta}). Additionally, clear reverberation signals have been detected in both lines, with a lag of $\sim$25 days (Figure \ref{fig:line_lc}), which may originate from the pre-existing broad-line region in AGN, as it follows the predicted radius-luminosity relation \citep{Du2018}.  
    
    \item We speculate that an rpTDE in AGN scenario may explain the observed dual flare. The UV/optical flare likely resulted from the collision of the TDE stream with the pre-existing accretion disk. This collision would deplete the inner region of the disk, causing the observed late-time UV dip \citep{Chan2020} (Figure \ref{fig:second_flare}). Furthermore, the corona may be partially destroyed due to the diminished gas supply from the inner disk \citep{Uttley2014} or strong outflows when accretion rate significantly increases \citep{Cao2023}, leading to a V-shaped pattern in the X-ray light curve. The absence of very broad emission lines could be explained by the re-accretion of stream debris by the accretion disk.
    
    \item In addition, our calculations suggest that the merger of a pair of sBHs within an accretion disk may also produce similar flares through BH-disk collisions. These flares could exhibit luminosities and durations comparable to those observed, making them challenging to distinguish from the stream-disk collision scenario in rpTDEs. 
\end{enumerate}
Assuming the same period, a third flare is expected within a few years, and observing more of these repeating flares could provide key evidence. The rpTDE can alter the star's orbit and structure with each encounter, in contrast to an sBH merger that tends to exhibit more stable periods and amplitudes. Well-designed monitoring of the next flare may yield decisive evidence, helping to distinguish between rpTDEs, sBH mergers, or other mechanisms.

In the future, time-domain surveys conducted by the Vera C. Rubin Observatory \citep[LSST,][]{Zeljko19} and the Wide Field Survey Telescope \citep[WFST,][]{WangTG23} are expected to identify more similar optical transients. When combined with parallel monitoring, e.g., the Einstein Probe X-ray observations \citep[EP,][]{Yuan22}, we anticipate being able to characterize multi-band properties, particularly lag information, through continuum RM \citep{Guo2025}. This approach is a promising tool for probing the unresolvable regions near SMBHs.

\vspace{5mm}
We thank the anonymous referee for helpful comments that improved the manuscript. We thank X.W., Cao, R.F. Shen, N. Jiang, and T.G. Wang for useful discussions. HXG acknowledges support from the National Key R\&D Program of China No. 2023YFA1607903 and 2022YFF0503402, National Natural Science Foundation of China (NSFC) No. 1247030223, Future Network Partner Program, CAS, No. 018GJHZ2022029FN, Overseas Center Platform Projects, CAS, No. 178GJHZ2023184MI. MFG is supported by the NSFC-12473019, the Shanghai Pilot Program for Basic Research-Chinese Academy of Science, Shanghai Branch (JCYJ-SHFY-2021-013), the National SKA Program of China (Grant No. 2022SKA0120102), the science research grants from the China Manned Space Project with No. CMSCSST-2021-A06. HCF acknowledges support from NSFC-12203096, Yunnan Fundamental Research Projects (grant NO. 202301AT070339). SSL acknowledges support from NSFC-12303022, Yunnan Fundamental Research Projects (grant No. 202301AT070358). Y.P.L. is supported in part by the NSFC-12373070 and NSFC-12192223, the Natural Science Foundation of Shanghai (grant NO. 23ZR1473700). WDZ and MHZ acknowledge NSFC-12333004, and support by the Strategic Priority Research Program of the Chinese Academy of Sciences, grant no. XDB0550200.

We are grateful for the high-SNR spectra observations under Director's Discretionary Time made with the GTC, installed at the Spanish Observatorio del Roque de los Muchachos of the Instituto de Astrof\'{i}sica de Canarias, on the island of La Palma. We acknowledge the support of the staff of the Lijiang 2.4m telescope. Funding for the telescope has been provided by Chinese Academy of Sciences and the People's Government of Yunnan Province. We acknowledge the support of the staff of the Xinglong 2.16m telescope. This work was partially Supported by the Open Project Program of the Key Laboratory of Optical Astronomy, National Astronomical Observatories, Chinese Academy of Sciences. We thank the rapid response of the Swift ToO observations. This work made use of data supplied by the UK Swift Science Data Centre at the University of Leicester \citep{Evans2007,Evans2009}. The National Radio Astronomy Observatory is a facility of the National Science Foundation operated under cooperative agreement by Associated Universities, Inc. Based on observations obtained with the Samuel Oschin Telescope 48-inch and the 60-inch Telescope at the Palomar Observatory as part of the ZTF project. ZTF is supported by the National Science Foundation under Grants No. AST-1440341 and AST-2034437 and a collaboration including current partners Caltech, IPAC, the Oskar Klein Center at Stockholm University, the University of Maryland, University of California, Berkeley, the University of Wisconsin at Milwaukee, University of Warwick, Ruhr University, Cornell University, Northwestern University and Drexel University. Operations are conducted by COO, IPAC, and UW.

\vspace{5mm}
\facilities{GTC, Swift, YAO:2.4m, Beijing:2.16m, VLA}

\software{Astropy \citep{2013A&A...558A..33A,2018AJ....156..123A},  
          Scipy \citep{2020SciPy}, 
          PyQSOFit \citep{Guo2018},
          PyCCF \citep{Sun2018},
          HEAsoft \citep{heasoft2014},
          CASA \citep{McMullin2007},
          PyRAF \citep{pyraf}} 
          
\bibliography{sample631}{}
\bibliographystyle{aasjournal}

\onecolumngrid
\appendix{}

\section{Radio Morphology and Spectrum}
\label{app:radio}
\begin{figure*}[htbp!]
    \centering
    \includegraphics[width=0.8\textwidth]{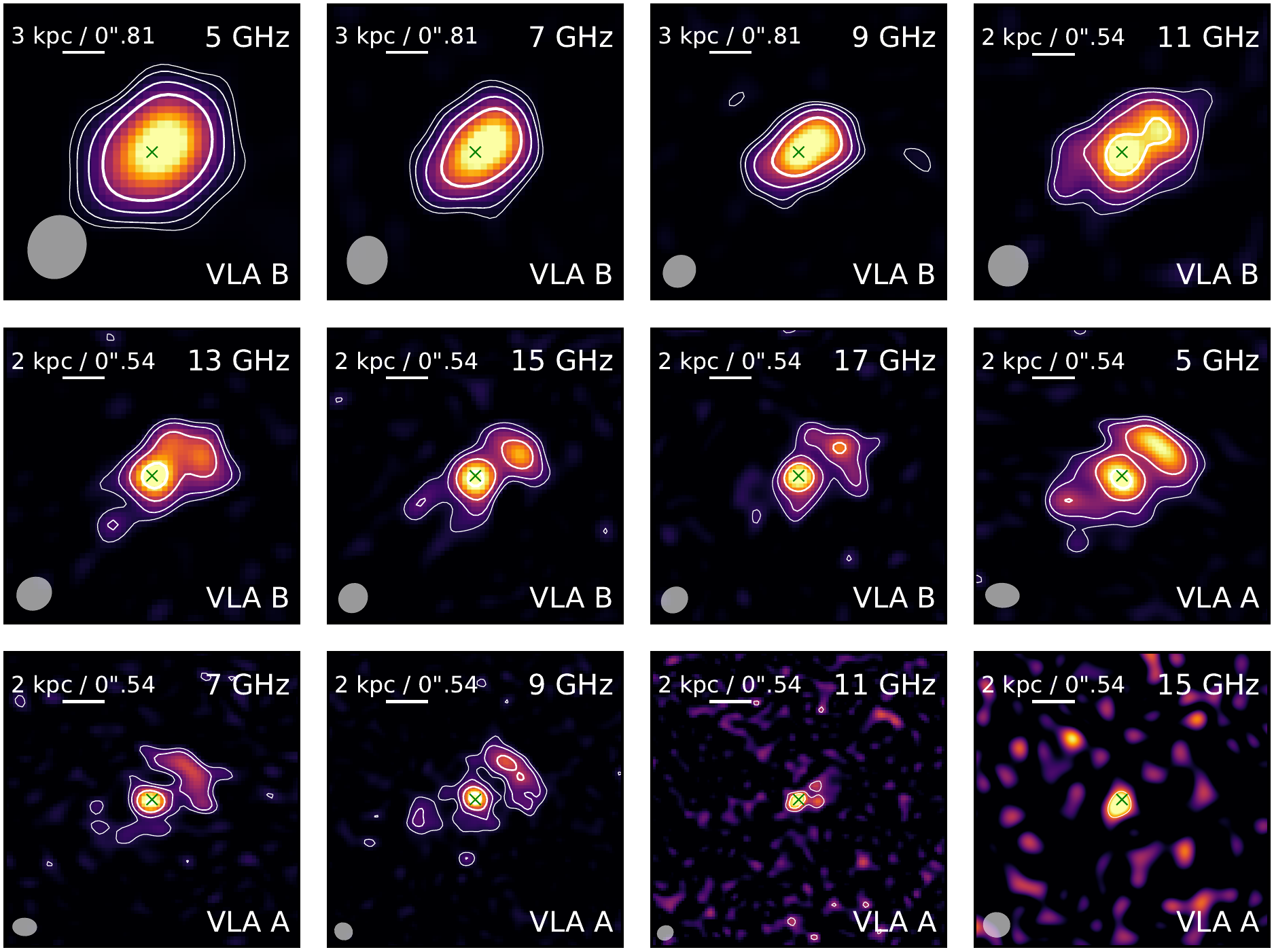}
    \caption{Naturally weighted CLEAN maps of VLA observations at 5$-$18 GHz in B-configuration and A-configuration. The green cross indicates the optical centroid from GAIA DR3. The grey ellipses in the bottom left corner of each panel represent the FWHM of the restoring beam.}
    \label{fig:vla} 
\end{figure*}

\begin{figure}
\begin{center}
    \includegraphics[width=0.6\textwidth]{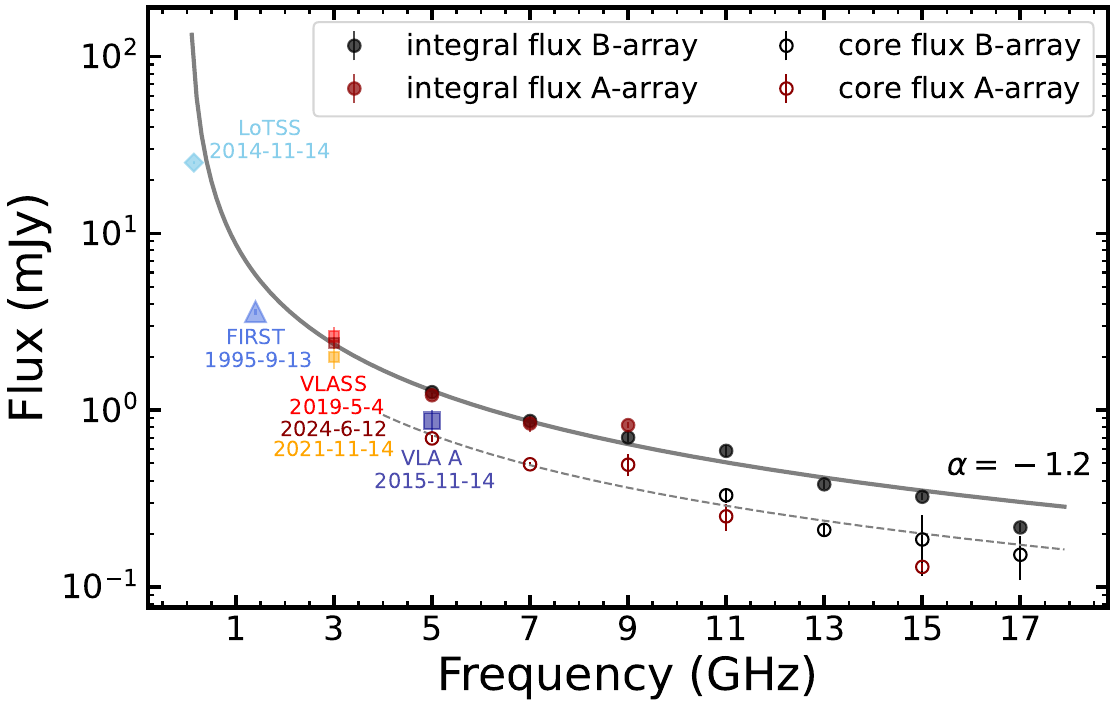}
    \caption{Radio spectrum of AT2021aeuk. The integrated flux is shown in solid circles, while the core flux is shown as open circles. The black and red symbols represent the flux in B-configuration and A-configuration, respectively. The black line indicates the best-fit power-law spectrum of the integrated flux with a spectral index $\alpha$ $-$1.2. The dashed line indicates the power-law fit to radio core with $\alpha$ of $-$1.1. The collected archival radio flux from VLA A-configuration, FIRST, VLASS, and LoTSS are shown in the figure.}
    \label{fig:radio_spectrum}
\end{center}
\end{figure}

The images from the VLA observations during the second flare at 4$-$18 GHz with B-configuration (+44 and +53 days) and A-configuration (+182 days) are shown in Figure~\ref{fig:vla}. Two epoch observations with B-configuration did not reveal significant flux variations and then were combined to make images with improved SNRs. The 5-GHz image with the A-configuration reveals two diffuse emission regions on either side of the nucleus, separated by a distance of 2 kpc. 
The radio spectra are shown in Figure~\ref{fig:radio_spectrum}. Our integrated VLA B-configuration flux measurements can be well-fitted by a power-law spectrum with an index of $\alpha = -1.2$. The integrated VLA A-configuration flux at 4-10 GHz don't show significant variations compared to B-configuration fluxes, thus are well matched with power-law spectrum. 

We retrieved archival radio detections from several surveys, including the Faint Images of the Radio Sky at Twenty-Centimeter \citep[FIRST,][]{Becker1995}, conducted in 1995; the LOFAR Two-meter Sky Survey \citep[LoTSS,][]{LoTSS2022}, in 2014; and three epochs of the VLA Sky Survey \citep[VLASS,][]{VLASS2020} in 2019, 2021, and 2024, corresponding to the period of the precursor flare, the first flare, and the second flare, respectively. The source was observed by the VLA in A-configuration at 5 GHz in 2015 \citep{Berton2018}. These archival radio fluxes align well with the power-law spectrum fitted to our VLA data, indicating most (if not all) radio emission is likely from AGN itself, rather than the fresh emission from the flare since no significant variation in radio emission after the outburst. The radio core is recognized with the brightest component at the locations matched with GAIA position. The spectrum of radio core can be well fitted with a single power-law with a steep spectral index of about $-1.1$ (see Figure~\ref{fig:radio_spectrum}).

This source was initially classified as radio-loud based on the 5 GHz flux density calculated from the FIRST 1.4 GHz flux assuming a flat radio spectrum. However, our VLA observations revealed less flux at 5 GHz than the expected value. Taking the flux densities of 1.0 mJy at 5 GHz and 0.26 mJy at $\rm 4400 \AA$ at source rest frame, the newly calculated radio loudness, ${\rm RL} = {S_{\rm 5 GHz}}/{S_{\rm 4400 \text{\AA}}} = 3.8$, identifying our target as a radio-quiet object \citep{Kellermann1989}. Radio loudness could be further reduced if taking the core flux from future VLBI observations if detected. The diffuse radio emission surrounding the nucleus, characterized by a steep spectrum, is likely dominated by star formation activities or AGN outflows at kpc scales \citep{Panessa2019}.

\section{Spectral Decomposition Results}
\begin{figure*}[htbp!]
\begin{center}
    \includegraphics[width=\textwidth]{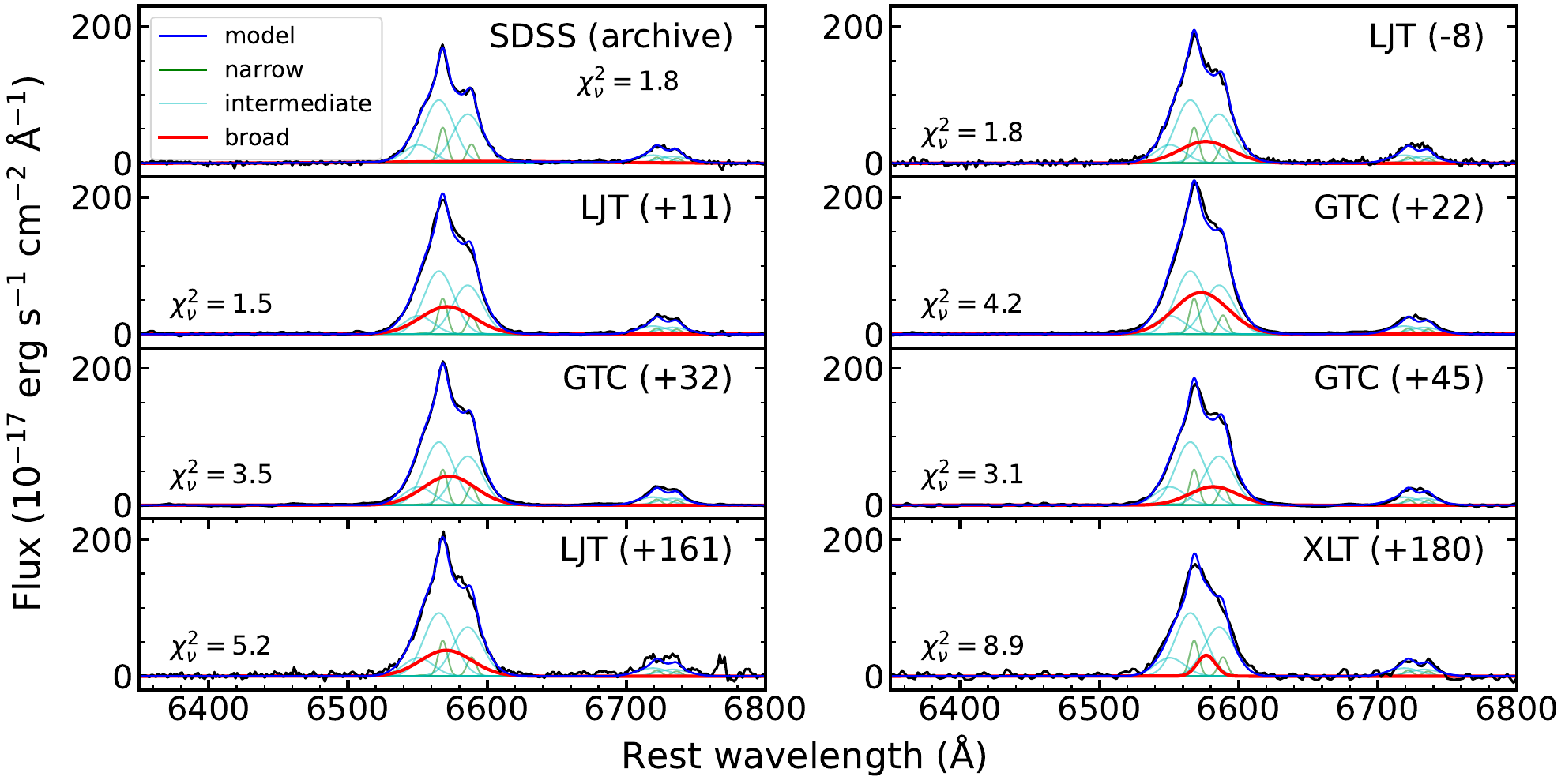}
    \caption{Spectral decomposition of the $\rm H\alpha$ complex. A narrow plus an intermediate component are used to fit each narrow line, and one additional broad component is used for $\rm H\alpha$.}
    \label{fig:h_alpha}
\end{center}
\end{figure*}

\begin{figure*}
\begin{center}
    \includegraphics[width=\textwidth]{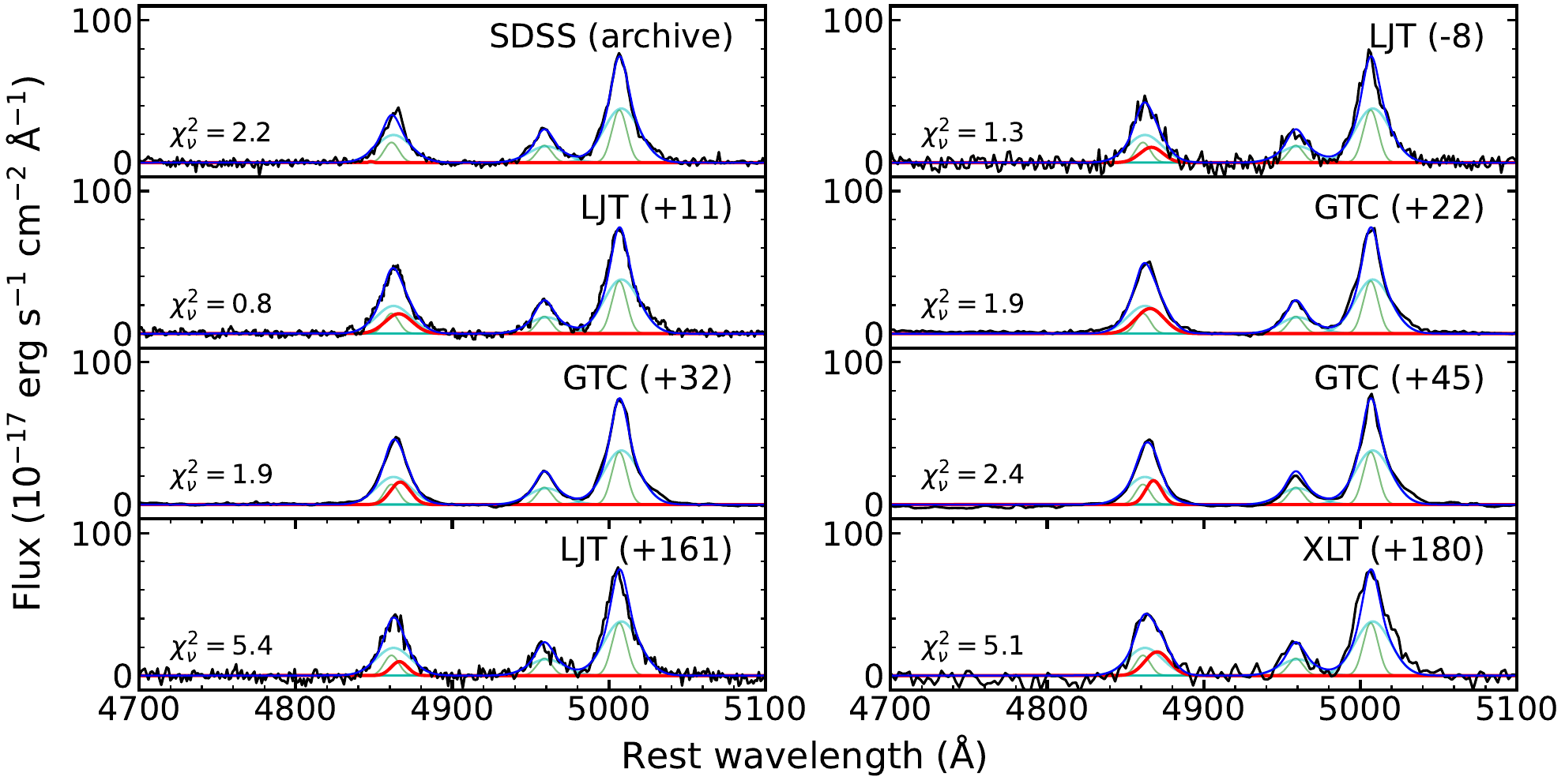}
    \caption{The same as Figure~\ref{fig:h_alpha}, but for $\rm H\beta$ complex.}
    \label{fig:h_beta}
\end{center}
\end{figure*}

We present the spectral decomposition results of H$\alpha$ and H$\beta$ in Figure~\ref{fig:h_alpha} and \ref{fig:h_beta}, respectively. In the archival SDSS spectrum, the emission lines are adequately fitted by narrow and intermediate components. During the second flare, a new broad component emerged in the spectra, exhibiting an FWHM of approximately 2000 $\rm km\ s^{-1}$ for H$\alpha$ and 1200 $\rm km\ s^{-1}$ for H$\beta$. 

\label{app:emission_line}

\section{Comparison with TDEs in quiescent galaxies and Other Nuclear Transients}
\label{app:compare}
\begin{figure*}
    \begin{center}
        \includegraphics[width=0.8\textwidth]{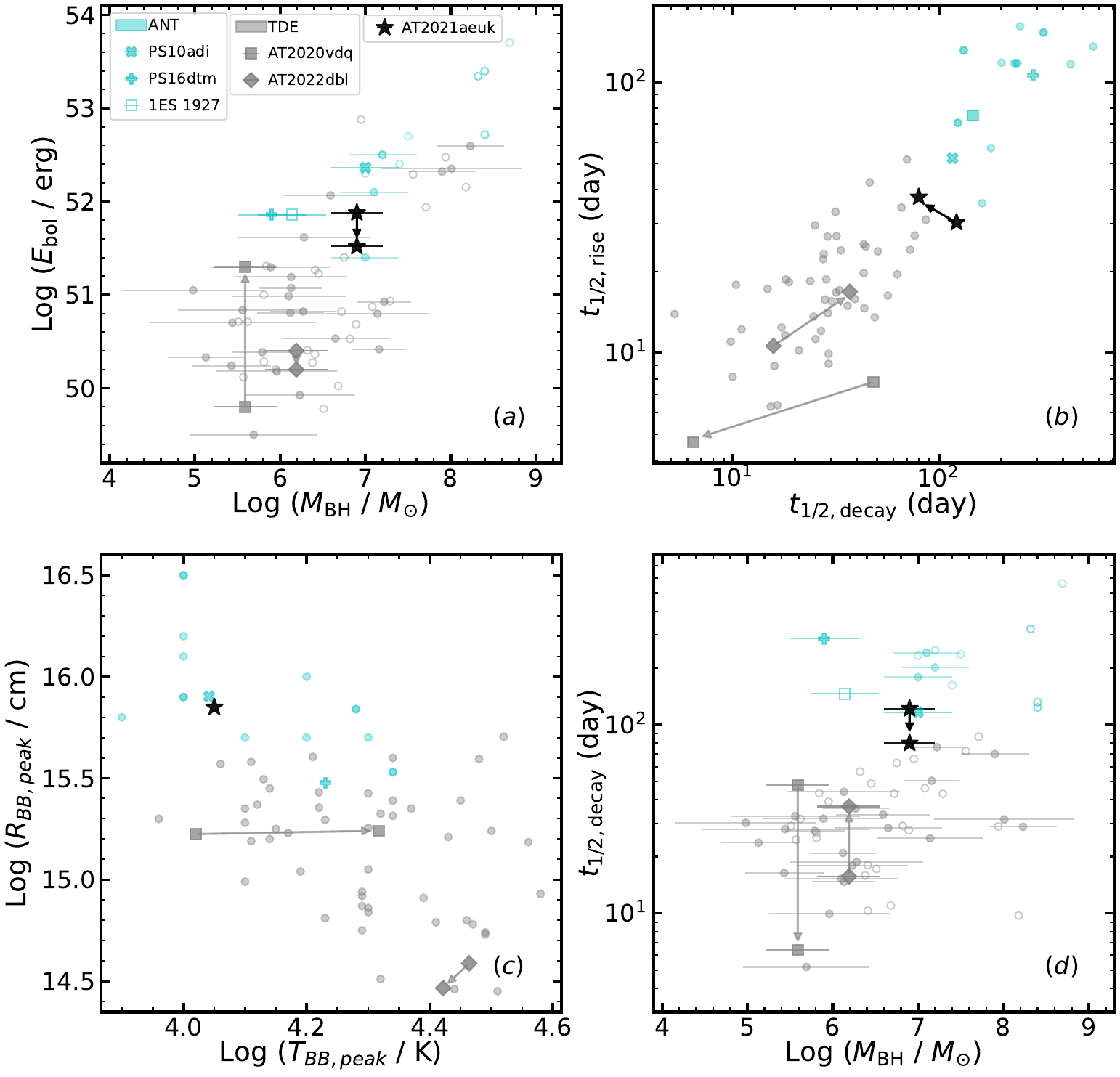}
    \caption{Comparison of AT2021aeuk with TDEs (gray), as well as ANTs and TDE in AGN candidates (cyan). Solid markers denote reliable BH mass while hollow markers, are considered less robust. For most light curve properties, the uncertainty from the fitting is negligible. For the rpTDE candidates, the arrows point from the first flare to the second flare. In the panel $c$, the $T_{\rm BB}$ and $R_{\rm BB}$ of AT2021aeuk is only available for the second flare.}
    \label{fig:properties}
\end{center}
\end{figure*}

We compared the light curve properties of AT2021aeuk with typical TDEs and other nuclear transients in Figure~\ref{fig:properties}. The comparison sample comprises two main categories: (1) systematically identified TDEs in quiescent galaxies by ZTF \citep{Hammerstein2023b, Yao2023}, along with two candidate rpTDEs in quiescient galaxies within the ZTF footprint, AT2020vdq and AT2022dbl; (2) the ambiguous nuclear transients (ANT) sample from \cite{Wiseman2024} and \cite{Hinkle2024}, as well as three TDE candidates in AGNs, PS10adi \citep{Kankare2017}, PS16dtm \citep{Blanchard2017, Petrushevska2023}, and 1ES 1927+654 \citep{Trakhtenbrot2019b, Ricci2020}.

For these transients, BH masses estimated by the $M_{\rm BH}-\sigma$ relation \citep{Gultekin2009, Kormendy2013} or the $R_{\rm BLR}-L$ relation \citep{Bentz2013} are indicated with solid markers in Figure~\ref{fig:properties}. Estimates based on the $M_{\rm BH}-M_{*}$ relation \citep{Greene2020} or light curve modeling \citep{Guillochon2018, Mockler2019} are considered relatively less reliable \citep{Hammerstein2023b} and are represented by hollow markers. As shown in Figure \ref{fig:properties}, the ANTs and TDE in AGN candidates generally exhibit higher bolometric energies, longer evolution times, lower blackbody temperatures, and larger blackbody radii compared to TDEs in quiescent galaxies with similar BH masses. The two flares of AT2021aeuk are located between the TDEs and ANTs.


\end{CJK}
\end{document}